\documentclass{aa}  
\usepackage{graphicx}
\usepackage{txfonts}
\usepackage{lipsum}

\usepackage{hyperref}
\usepackage{xcolor}

\newcommand{\postref}[1]{{#1}}

\newcommand{\postreftwo}[1]{{#1}}

\newcommand{\postlang}[1]{{#1}}

\def\deg{\ifmmode^\circ\else$^\circ$\fi}
\def\pdeg{\ifmmode $\setbox0=\hbox{$^{\circ}$}\rlap{\hskip.11\wd0 .}$^{\circ}
          \else \setbox0=\hbox{$^{\circ}$}\rlap{\hskip.11\wd0 .}$^{\circ}$\fi}
\def\arcs{\ifmmode {^{\scriptstyle\prime\prime}}
          \else $^{\scriptstyle\prime\prime}$\fi}
\def\arcm{\ifmmode {^{\scriptstyle\prime}}
          \else $^{\scriptstyle\prime}$\fi}
\newdimen\sa  \newdimen\sb
\def\parcs{\sa=.07em \sb=.03em
     \ifmmode \hbox{\rlap{.}}^{\scriptstyle\prime\kern -\sb\prime}\hbox{\kern -\sa}
     \else \rlap{.}$^{\scriptstyle\prime\kern -\sb\prime}$\kern -\sa\fi}
\def\parcm{\sa=.08em \sb=.03em
     \ifmmode \hbox{\rlap{.}\kern\sa}^{\scriptstyle\prime}\hbox{\kern-\sb}
     \else \rlap{.}\kern\sa$^{\scriptstyle\prime}$\kern-\sb\fi}
\def\kps{km\,s$^{-1}$}
\def\micron{$\mu$m}

\def\msun{M$_{\odot}$}

\defcitealias{pabst2020}{P20} 
                                
\begin{document}

\title{The \postlang{N}eutral \postlang{A}tomic \postlang{H}ydrogen in the solar neighborhood (NeAtHood) project}
\subtitle{I. Ghost in the shell: Neutral atomic hydrogen in the extended Orion nebula}
\titlerunning{Neutral atomic hydrogen in the extended Orion nebula}


   \author{J.~D.~Soler\inst{1}\fnmsep\thanks{ Corresponding author; \email{juandiegosolerp@gmail.com}}
        \and H.~Beuther\inst{2}
        \and S.~C.~O.~Glover\inst{3}
        \and R.~S.~Klessen\inst{3,4}
        \and J.~Ott\inst{5}
        \and M.~Rugel\inst{6}
        \and J. W. Teh\inst{3} 
        \and S.~E.~Clark\inst{7,8}
        \and P.~Goldsmith\inst{9}
        \and A.~Hacar\inst{1}
        \and A.~Socci\inst{1}
        \and M.~Heyer\inst{10}
        \and M.-Y.~Lee\inst{11}
        \and C. E. Murray\inst{12}
        \and D.~Seifried\inst{13}
        \and S. Walch\inst{13}
        \and B.~Godard\inst{14,15}
        \and M.-A.~Miville-Desch\^{e}nes\inst{15}
        }

   \institute{University of Vienna, Department of Astrophysics, T\"{u}rkenschanzstra{\ss}e 17, 1180 Wien, Austria \and 
   Max-Planck Institute for Astronomy, K\"{o}nigstuhl 17, 69117 Heidelberg, Germany \and
   Universit\"{a}t Heidelberg, Zentrum f\"{u}r Astronomie, Institut f\"{u}r Theoretische Astrophysik, Albert-Ueberle-Str. 2, 69120 Heidelberg, Germany \and
   Universit\"{a}t Heidelberg, Interdisziplin\"{a}res Zentrum f\"{u}r Wissenschaftliches Rechnen, Im Neuenheimer Feld 225, 69120 Heidelberg, Germany \and 
   National Radio Astronomy Observatory, PO Box O, 1003 Lopezville Road, Socorro, NM 87801, USA \and
   Deutsches Zentrum for Astrophysics, Postplatz 1, 02826 G\"{o}rlitz, Germany \and
   Department of Physics, Stanford University, Stanford, CA 94305, USA \and 
   Kavli Institute for Particle Astrophysics \& Cosmology, P.O. Box 2450, Stanford University, Stanford, CA 94305, USA \and
   Jet Propulsion Laboratory, California Institute of Technology, 4800 Oak Grove Drive, Pasadena CA, 91109, USA \and
   Department of Astronomy, University of Massachusetts, Amherst, MA 01003-9305, USA \and 
   Korea Astronomy \& Space Science Institute,776 Daedeok-daero, Yuseong-gu, Daejeon 34055, Republic of Korea \and
   Department of Physics \& Astronomy, Johns Hopkins University, 3400 N. Charles Street, Baltimore, MD 21218, USA \and
   Universit\"{a}t zu K\"{o}ln, I. Physikalisches Institut, Z\"{u}lpicher Str. 77, D-50937 K\"{o}ln, Germany \and
   Observatoire de Paris, PSL University, Sorbonne Universit\'{e}, LERMA, 75014 Paris, France \and
   Laboratoire de Physique de l’\'{E}cole Normale Sup\'{e}rieure, ENS, Université PSL, CNRS, Sorbonne Universit\'{e}, Universit\'{e} de Paris
}
   \date{Received: 02FEB2026. Accepted: 27APR2026}

  \abstract{
  The Orion nebula is the nearest site of ongoing and recent high-mass star formation. It is a unique laboratory for studying the mass, energy, and momentum input from high-mass stars.
  We present 21-centimeter emission line observations that resolve for the first time the neutral atomic hydrogen (H{\sc i}) gas in the extended Orion nebula (EON) at a resolution of one arcminute, which \postref{corresponds} to a physical scale of 0.12 parsecs at the standard distance to the region.
  Our H{\sc i} emission maps reveal an expanding shell that matches the EON contours delineated by recent observations of ionized carbon ([C{\sc ii}]) line emission.
  However, our combination of single-dish and interferometric H{\sc i} observations suggests 100 solar masses of material for the front hemisphere of the shell, which is lower by roughly a factor of ten than the mass inferred from [C{\sc ii}] observations.
  This discrepancy suggests that the mass of the nearest wind-blown bubble has been overestimated, although we do not rule out the possibility that a significant amount of molecular hydrogen (H$_{2}$) in the shell may account for part of the difference.
  Our extended \postlang{21 cm} line maps also reveal uncharted structures in and around the EON.
  They include a probable secondary bubble and a linear protrusion extending roughly four parsecs from the shell boundary.
  Our results illustrate the potential of H{\sc i} interferometric observations to elucidate key aspects of the multiphase structure of star-forming regions and their connection to their surroundings.
  }

\maketitle
\nolinenumbers

\section{Introduction}

Neutral atomic hydrogen (H{\sc i}) is a fundamental tracer of the cycling of matter and energy in galaxies \citep{oort1958,kalberla2009,mcclure-griffiths2023}.
H{\sc i} comprises roughly two-thirds of the gas in the Milky Way and traces the cold pre-molecular state before star formation and the warm diffuse interstellar medium before and after star formation \citep{klessen2016}.
However, H{\sc i} emission maps of the nearest star-forming regions have been limited in extension and resolution.
We present the first result from the project called neutral atomic hydrogen in the solar neighborhood (NeAtHood), which aims to produce and study arcminute-resolution H{\sc i} maps obtained with the Karl G. Jansky Very Large Array (VLA).
The target of this study is one of the brightest and most recognizable objects in the night sky: the Orion nebula.

\begin{figure}[h!]
\centering{
\includegraphics[width=0.48\textwidth,angle=0,origin=c]{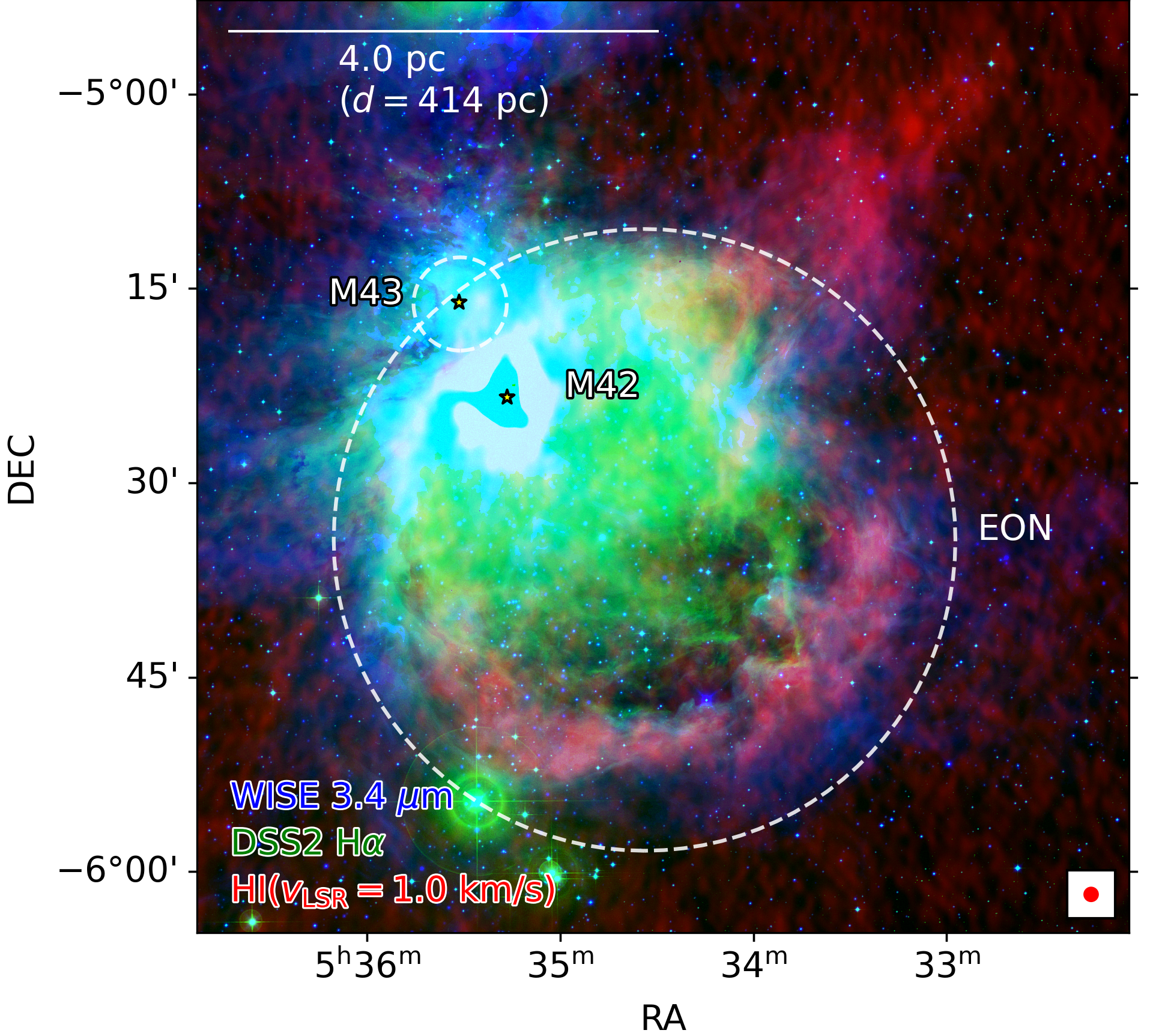}
\caption{Extended Orion nebula shell sampled by H{\sc i} emission at $v_{\rm LSR}$\,$=$\,1.0\,\kps\ from the combined VLA and FAST observations (shown in red), H$\alpha$ emission from the European Southern Observatory Digitized Sky Survey (shown in green), and 3.4-$\mu$m emission registered by the Wide-field Infrared Survey Explorer (WISE) satellite  (shown in blue). 
The dashed white circles indicate the locations of the EON and M43 shells.
The yellow stars show the position of their presumed progenitors, O7V-type star $\theta^{1}$\,Ori\,C and B3V/IV-type star $\nu$ Ori. 
The effective angular resolution of the H{\sc i} \postlang{21 cm} observations is indicated by the red disk in the lower right corner.
}\label{fig:EONhiANDwise}
}
\end{figure}

The Orion nebula (Messier 42; M42; NGC\,1976) is the only ionized nebula visible to the naked eye and has been the subject of astronomical research for at least four centuries \citep{herczeg1998}.
There is no \postlang{standard} division scheme for M42, but astronomers commonly {split} it into several well-defined subregions based on the wavelength and scientific focus of the observations, as illustrated in Fig.~\ref{fig:EONhiANDwise}.
The bright inner \postlang{zone} is often referred to as the Huygens region and contains the Trapezium cluster. This is the nearest region of recent high-mass star formation\postlang{,} a cornerstone of studies of stellar clusters and early stellar evolution \citep{hillenbrand1997,odell2001,bally2008}.
The brightest ionized gas associated with the Trapezium stars lies at approximately 414\,pc from the Sun \citep{menten2007}.

\begin{figure}[h!]
\centering{
\includegraphics[width=0.5\textwidth,angle=0,origin=c]{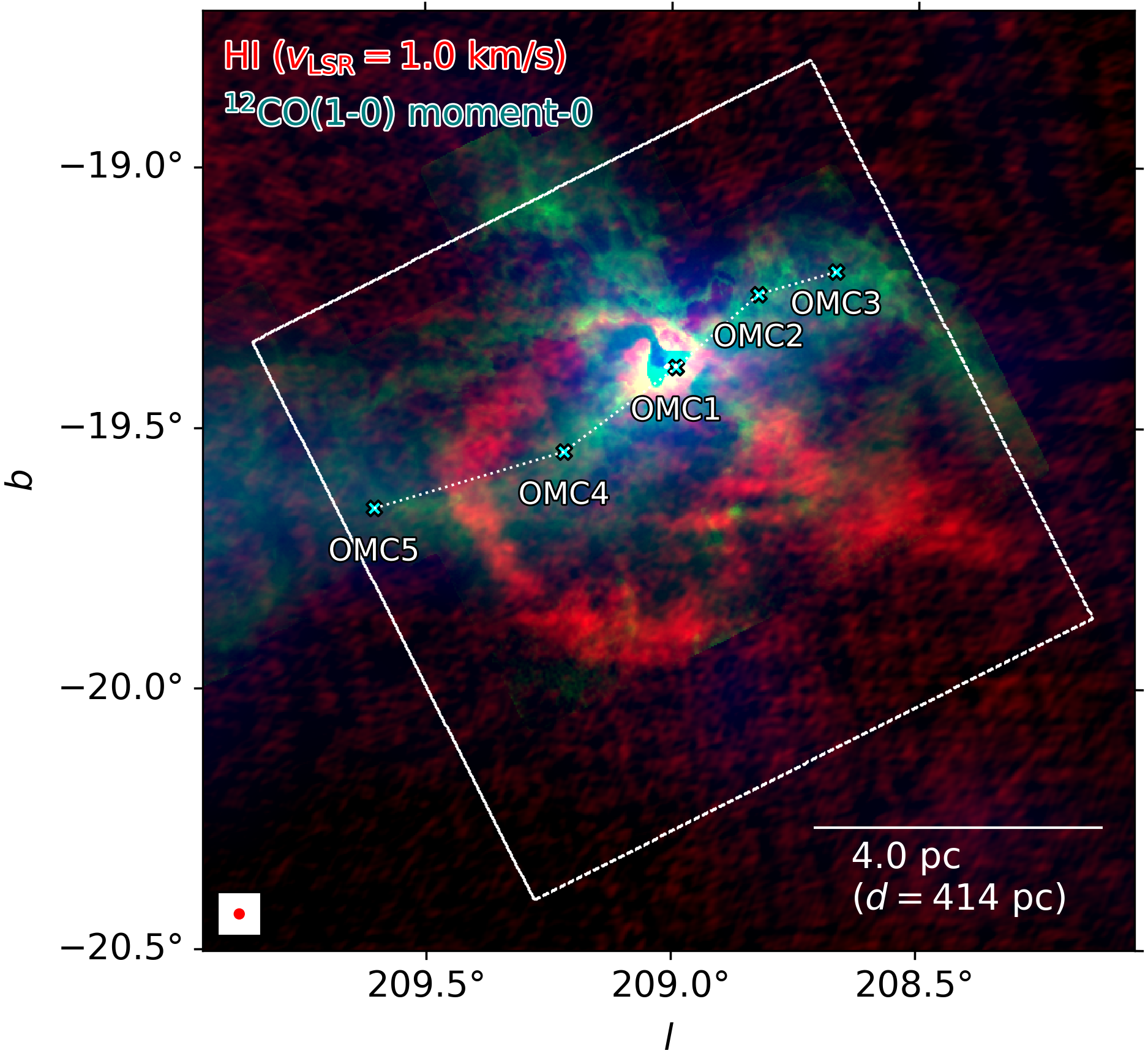}
\caption{Extended Orion nebula and its surroundings in Galactic coordinates and in the context of the  \postlang{OMC}, shown in the $^{12}$CO$(1\,$\,$\rightarrow$\,$0)$ line emission from the Orion CARMA survey \citep{kong2018} and the \cite{dame2001} survey.
The crosses indicate the central positions of different OMC components identified in the structure known as the integral-shaped filament, roughly indicated by the dotted lines. 
The dashed square corresponds to the region shown in Fig.~\ref{fig:EONhiANDwise}.
}\label{fig:EONhiANDco}
}
\end{figure}

M42 is associated with the Orion molecular cloud (OMC) system, which is conspicuously revealed by carbon monoxide emission \citep[][]{wilson2005,berne2014,kong2018}, as shown in Fig.~\ref{fig:EONhiANDco}.
The OMC has been extensively sampled across wavelengths, from gamma-ray \citep{ackermann2012} to the radio \citep{subrahmanyan2001}, and in continuum and line emission \citep[see, for example,][]{megeath2012,polychroni2013,shimajiri2015}.
One of the most conspicuous features in the gas around the OMC is a large hemispherical shell usually called the extended Orion nebula (EON) on the observer’s side of the molecular cloud \citep[][]{gudel2008,odell2017}. 
Our data \postlang{resolve} it for the \postlang{first} time in H{\sc i} emission.

The origin of the EON shell is usually attributed to the hot plasma bubble created by the winds from the O7V star $\theta^{1}$\,Ori\,C \citep{abel2004,odell2017,pabst2019}.
The hot plasma in the interior of the cavity and its corresponding expanding shell have been identified in observations of soft X-ray emission \citep{gudel2008} and ionized carbon fine-structure ([C{\sc ii}]) line emission at 158\,\micron\ 
\citep[][from here on \citetalias{pabst2020}]{pabst2020}, respectively.
The [C{\sc ii}] emission traces a combination of the material in photodissociation regions (PDRs), CO-dark molecular gas, and in the cold neutral medium (CNM).
The H{\sc i} observations necessary to separate the contributions from each of these components in the EON shell have been critically limited by angular resolution so far.

Observations of the H{\sc i} \postlang{21 cm} line toward Orion span more than seven decades \citep{bok1955,menon1958,gordon1970,heilesANDhabing1974}.
\cite{wannier1983} made the first attempt to establish a detailed correlation between the molecular material traced by the carbon monoxide (CO) emission in the OMC and the H{\sc i} emission in a strip sampled with the 305-m reflector of the Arecibo Observatory. 
\cite{chromey1989} presented a study of the H{\sc i} emission in a 10$^{\circ}$\,$\times$\,18$^{\circ}$ region around the molecular clouds at 20\arcmin\ resolution using the National Radio Astronomy Observatory (NRAO) 140-foot (43-meter) telescope, showing that the atomic gas mass around M42 is comparable to that in dense molecular clouds, around 2\,$\times$\,10$^{5}$\,M$_{\odot}$.
\cite{green1991} employed the Dominion Radio Astrophysical Observatory (DRAO) 26-m telescope to sample H{\sc i} at a resolution of 36\arcmin\ in a 28.5\pdeg\,$\times$\,28\pdeg5 region around Orion, reporting large-scale features associated with the Orion-Eridanus superbubble \citep{green1993}. 
\cite{vanderWerf2013} employed the VLA C- and B-array configurations to sample M42 in a region of approximately 16\arcmin\,$\times$\,16\arcmin\ covering the position of the Trapezium stars, the Orion Bar, and {M43} at 7\parcs2\,$\times$\,5\parcs7 resolution.
However, until very recently, the observations with highest angular resolution H{\sc i} of the extended region around M42 came from the Parkes Galactic All Sky H{\sc i} Survey \citep[GASS;] []{mcclure-griffiths2009,kalberla2010}, at 16\parcm2 FWHM.

We present a combination of observations from the VLA and the 500-m Aperture Spherical Radio Telescope (FAST) that result in the first 1\arcmin-resolution map of H{\sc i} \postlang{21 cm} emission toward the EON shell, shown in Fig.~\ref{fig:EONhiANDwise}.
This angular resolution for the first time matches the resolution of tens of arcseconds of [C{\sc ii}] and carbon monoxide (CO) line emission toward this region \citep{kong2018,pabst2019}.
We describe the main features revealed by this new dataset and compare them with the conclusions derived from previous observations, with particular focus on [C{\sc ii}].
In Section~\ref{sec:data} we describe the data and the methods we employed to construct the H{\sc i} maps.
Section~\ref{sec:results} presents the physical properties derived from the H{\sc i} observations.
We discuss the implications of our results in Sec.~\ref{sec:discussion} and present our conclusions in Sec.~\ref{sec:conclusions}.
We reserve the technical details of the interferometric data processing for two dedicated appendices.
Appendix \ref{app:VLA} presents the flagging, calibration, and imaging procedures.
Appendix \ref{app:FAST} details the combination with single-dish observations and the quality-control tests applied to the final data product. 
Finally, Appendix \ref{app:NGC1977} presents the maps of the H{\sc ii} region NGC 1977, which is also covered in our interferometric mosaics, but whose detailed study is beyond the scope of this work.

\section{Data}\label{sec:data}

\begin{figure*}[h!]
\includegraphics[width=0.99\textwidth,angle=0,origin=c]{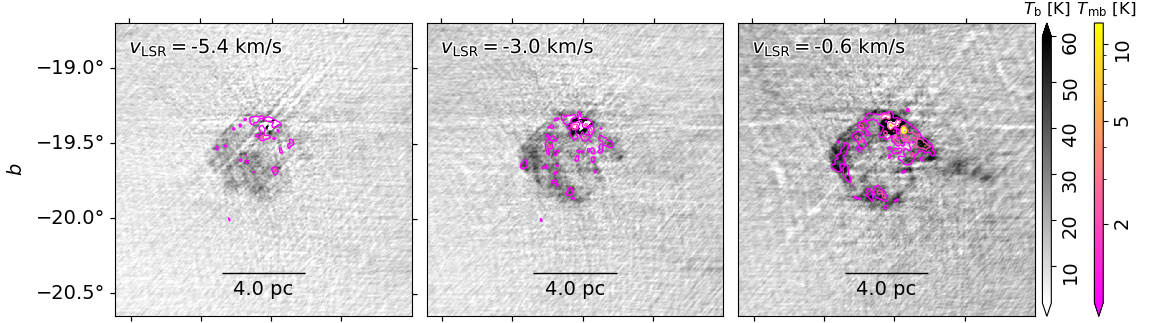}
\includegraphics[width=0.99\textwidth,angle=0,origin=c]{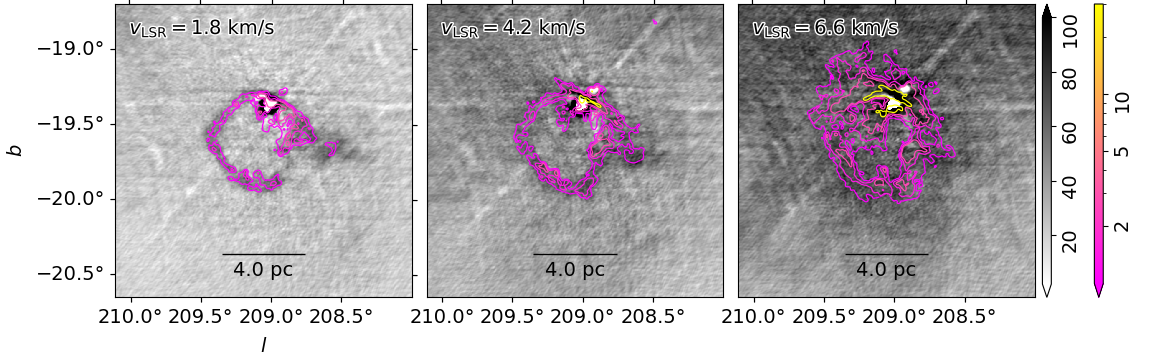}
\caption{Selection of H{\sc i} \postlang{21 cm} line emission maps from the combination of VLA-D and FAST data (grayscale) and [C{\sc ii}] line emission (contours) for 0.4-\kps-wide velocity channels around the indicated $v_{\rm LSR}$.
The [C{\sc ii}] line emission contours correspond to main-beam brightness temperatures of $T_{\rm mb}$\,$=$\,2, 5, 10, and 30\,K.
The minimum values of the two maps are set to the observation noise levels.
}
\label{fig:multivpanels}
\end{figure*}

\subsection{VLA}

The primary dataset for this work comes from the National Radio Astronomy Observatory (NRAO)'s  VLA project Orion Dynamics of Ionized and Neutral gas (ODIN; 19B-181; PI: J.~D.~Soler).
The ODIN survey design and observation strategy are based on the legacy of The HI/OH/Recombination line survey of the inner Milky Way \citep[THOR;][]{beuther2016,wang2020hi}.
The data were acquired in 1\pdeg25\,$\times$\,2\pdeg5 tiles covered with 5\,$\times$\,9 pointings of the VLA D-array.
For this work, we employed two tiles centered on $[l,b]$\,$=$\,[208\pdeg75,$-$19\pdeg5] and [210\pdeg00,$-$19\pdeg5] observed in two scheduling blocks executed in December 2019.
Following the THOR correlator setup, we observed the radio continuum in eight bands spanning 1 and 2\,GHz in total intensity and polarization, as well as four OH lines, 19 H$n\alpha$ radio recombination lines, and the H{\sc i} \postlang{21 cm} line in 0.4-\kps wide velocity channels. 
The D configuration covers baselines {raging from} 39.9 {to} 1031.2\,m, with an average separation of 379.6\,m.
The resulting synthesized beam is 70\parcs4 by 51\parcs6, with a position angle of $-$4\pdeg2, corresponding to an effective angular resolution of 60\parcs3 FWHM.

We employed the Common Astronomy Software Applications (CASA) software version 6.6.6.17 for the data reduction.
The radio frequency interference (RFI) was flagged using the VLA pipeline automated routine.
Flux, bandpass, and polarization calibrations were performed using quasar 3C\,48 as reference source.
This strong calibrator is not devoid of H{\sc i} contamination \citep[][]{murray2015}, so we performed a polynomial interpolation of its spectra using CASA's {\tt bandpass} procedure with parameter {\tt bandtype}\,$=$\,{\tt BPOLY}.
A flare started in this calibration source in 2018, which {might} shift {the} flux calibration by a few percent.
Quasar J0532+0732 (QSO B0529+075) was used as a complex-gain calibrator. 
Further details on the data reduction procedure are presented in Appendix\ref{app:VLA}.

We imaged the H{\sc i} emission in both mosaics using the CASA {\tt tclean} routine.
Following the extensive test performed with the THOR H{\sc i} observations \citep{beuther2016}, we chose the {\tt mosaic} gridding algorithm and the H\"{o}gbom deconvolution procedure.
We employed Briggs weighting with a robustness parameter of 0.5.
We set the cleaning threshold at 50\,mJy/beam and {\tt gain}\,$=$\,0.1.
The main challenge in imaging this region is the bright radio continuum {emission} from M42, which we mitigated by applying a bright-pixel limit {\tt bplim}\,$=$\,0.2.
Exploration of alternative parameters yielded less satisfactory images. 
We therefore report the results obtained with this combination.

\subsection{FAST}

We combined our interferometric data with the observations in the Commensal Radio Astronomy FAST Survey \citep[CRAFTS,][]{li2018}, a FAST legacy survey.
CRAFTS is an ongoing program to sample the sky in the declination range $-14$\deg\,$<$\,$\delta$\,$<$\,66\deg, covering over 20,000 square degrees.
To achieve this observation range, FAST was fully illuminated at zenith angles up to 26\pdeg4 and partially illuminated at zenith angles up to 40\deg\ \citep{jiang2020}.
CRAFTS was conducted as a two-pass drift-scan survey by the FAST L-band Array of 19-beam (FLAN) receiver, which was rotated by 23\pdeg4 to achieve a higher-than-Nyquist sampling while drifting \citep{li2018}, thus minimizing radio frequency interference (RFI) and system temperature variations introduced when radiation from the surrounding mountain peaks enters the nearer sidelobes \citep{zhang2019}.

CRAFTS has thus far observed approximately 20\% of the planned sky coverage. 
We used observations from the CRAFTS H{\sc i} Narrow All-sky (CH{\sc i}NA) survey (Kr\^{c}o et al., in prep.).
The first data release (DR1) of CH{\sc i}NA, containing around 4,500\,deg$^{2}$ of calibrated publication-quality images in 10\deg\,$\times$\,10\deg\ Stokes $I$ cubes available in the H{\sc i}Verse platform\footnote{\url{https://hiverse.alkaidos.cn/}}. 
The published cubes cover the range $-600$\,$<$\,$v_{\rm LSR}$\,$<$\,600\,\kps in 0.2\,\kps\ channels with a nonuniform sensitivity of around 0.17\,K or better. 
The original FAST beam size is 2\parcm9, but the effective beam size of CH{\sc i}NA observations is 4\parcm0 FWHM.

We employed the CH{\sc i}NA cube centered on $\delta$\,$=$\,$-10$\deg\ and right ascension $\alpha$\,$=$\,85\deg.
We removed additional RFI features in the velocity range of interest by applying a baseline correction by subtracting a fourth-order polynomial fit to the data in the $|v_{\rm LSR}|$\,$>$\,100\,\kps.
FAST does not provide reliable absorption measurements toward the Trapezium cluster, so we excluded that region from our analysis and masked its surroundings in the combined data. 

\subsection{VLA and FAST combination}

We combined the single-dish and interferometric data using the Fourier-transform-based feathering algorithm, implemented in CASA through the {\tt feather} function.
For this purpose, we projected the FAST data onto the same grid as our VLA mosaic using the {\tt reproject} package \citep{robitaille2020}.
We smoothed and resampled the spectra to the 0.4-\kps spectral resolution of the VLA observations employing the {\tt spectral\_cube} package in {\tt astropy}.
The scale factor applied by the algorithm to the single-dish image was kept at the default value ({\tt sdfactor}\,$=$\,1.0).
An example of the output velocity-channel maps is shown in Fig.~\ref{fig:multivpanels}.
The estimated noise of the combined data is approximately 2.0\,K, as further detailed in Appendix\ref{app:FAST}.

The lack of FAST data toward the Orion nebula cluster (ONC) results in a ring-like feature around that region.
This feature arises from the combination of the negative signal produced by H{\sc i} absorption toward the strong continuum source in the interferometric data and the absence of single-dish data.
We avoided this spurious characteristic by masking a region with a radius of 12\arcmin around the central cluster location ($[l,b]$\,$=$\,$[$209\pdeg06,$-$19\pdeg48$]$).
Our subject of interest is the tens-of-arcminute EON bubble and its surroundings, and therefore, it is not critical for our results \postlang{of our analysis}.

\section{Results}\label{sec:results}

\begin{figure*}[h!]
\sidecaption
\includegraphics[width=0.35\textwidth,angle=0,origin=c]{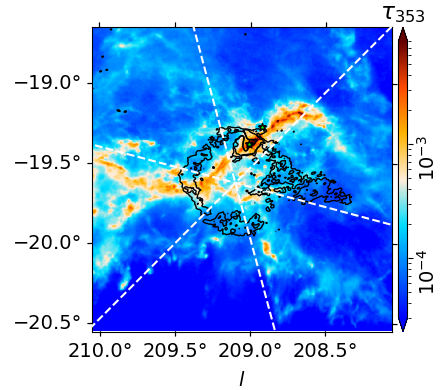}
\includegraphics[width=0.35\textwidth,angle=0,origin=c]{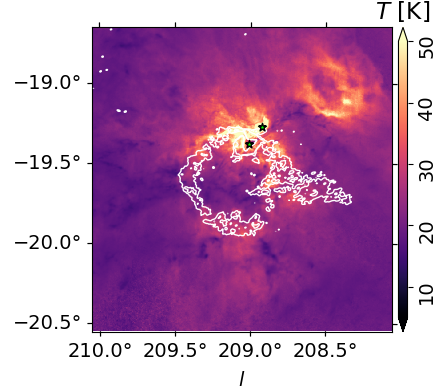}
\caption{Contours of H{\sc i} line emission at $v_{\rm LSR}$\,$=$\,1.0\,\kps\ overlaid on the dust optical depth at 353\,GHz ($\tau_{353}$) and mean dust temperature estimated from the combination of {\it Planck} and {\it Herschel} observations in \cite{lombardi2014}. 
The stars mark the positions of $\theta^{1}$ Ori C and HD37061.
The dashed white lines indicate the center of the expanding shell and the direction of the radial profiles presented in Fig.~\ref{fig:radialprofiles}.
}\label{fig:dusttauandT}
\end{figure*}

Spectroscopic images of H{\sc i} emission at different velocities provide valuable insights into the dynamics of the atomic gas, its interactions with the molecular gas component, and the {expansion} driven by stellar winds from young high-mass stars.
Figure~\ref{fig:multivpanels} presents a selection of H{\sc i} emission maps obtained from our combination of the VLA and FAST data.
The most prominent feature in the H{\sc i} emission across $-10$\,$<$\,$v_{\rm LSR}$\,$<$\,2\,\kps\ is the circular structure that we identify as the EON shell.
For $v_{\rm LSR}$\,$<$\,$-2$\,\kps, this feature is filled with emission, which we interpret as the material on the observer's side of the EON shell.
For $v_{\rm LSR}$\,$>$\,$1$\,\kps, the shell is accompanied by extended H{\sc i} emission (lower right panel of Fig.~\ref{fig:multivpanels}), which is most likely associated with the Orion A molecular cloud, which is clearly distinguishable in the CO emission in the 2\,$<$\,$v_{\rm LSR}$\,$<$\,12\,\kps\ range \citep{wilson2005}.

Compared with the structures revealed by the [C{\sc ii}] emission, also shown in Fig.~\ref{fig:multivpanels}, the H{\sc i} emission is more extended throughout the front of the shell.
The H{\sc i} emission also displays an elongated protrusion on the west side of the bubble (right side of Fig.~\ref{fig:multivpanels}).
The base of this protrusion corresponds to the structure identified in [C{\sc ii}] emission in \cite{kavak2022protrusion}, but is clearly larger in the \postlang{21 cm} emission, as shown in the bottom left panel of Fig.~\ref{fig:multivpanels}.
This structure, which we refer to as the EON protrusion, has no evident counterpart in CO emission, as shown in Fig.~\ref{fig:EONhiANDco}, or in the dust column density and temperature, as presented in Fig.~\ref{fig:dusttauandT}.

\begin{figure*}[h!]
\centering{
\includegraphics[width=0.33\textwidth,angle=0,origin=c]{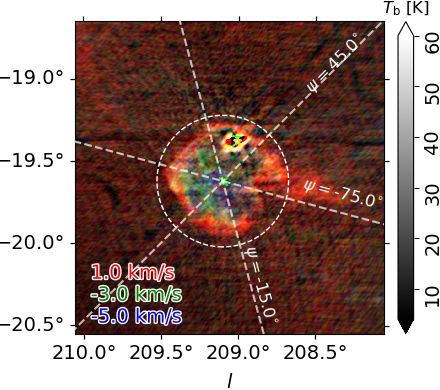}
\includegraphics[width=0.33\textwidth,angle=0,origin=c]{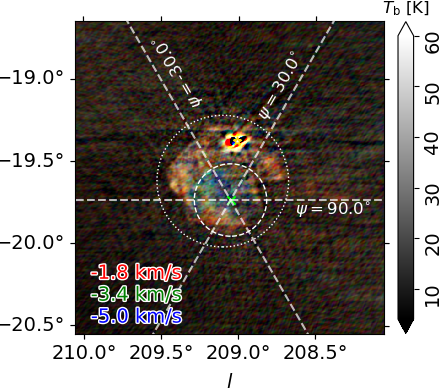}
\includegraphics[width=0.33\textwidth,angle=0,origin=c]{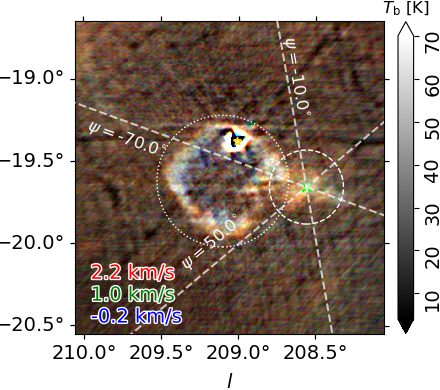}
}
\caption{H{\sc i} emission toward three regions of interest in and around the EON bubble.
The colors represent the emission in {0.4-\kps wide channels centered on the indicated velocities}.
The dashed white lines indicate the direction of the radial profiles presented in Fig.~\ref{fig:radialprofiles}.
{The orientation angle $\psi$ identifies the profile orientation using the IAU polarization convention, i.e., measured with respect to the north Galactic pole with positive angles measured clockwise.
The dashed circles correspond to locations of the guidelines in the position-velocity diagram presented in Fig.~\ref{fig:lvEONbubble}, Fig.~\ref{fig:lvSECbubble}, and Fig.~\ref{fig:lvEONchimney}.
For scale comparison, we included the dashed circle in the left panel as dotted circles in the middle and right panels.}
}\label{fig:bubbleRGB}
\end{figure*}

The H{\sc i} absorption against the radio continuum emission in the Huygens region is evident across velocity channels in the VLA data, as shown {in} Appendix~\ref{app:VLA}.
In the range $v_{\rm LSR}$\,$\gtrsim$\,$3$\,\kps, for which the H{\sc i} emission level increases outside of the bubble, the H{\sc i} absorption also reveals M43 in the upper right corner of the EON shell.
For these velocities, the \postlang{21 cm} emission displays wispy structures in the bulk of the emission surrounding the bubble.
These are not the result of the interferometric image pattern, as they are also discernible in the FAST data; we reserve a detailed study for a subsequent work.
In what follows, we describe the physical properties of the EON shell and other related structures revealed by the H{\sc i} emission.

\begin{table}[ht!]
\caption{\label{table:EONprops}Properties of the EON shell front hemisphere derived from H{\sc i} emission}
\centering
\begin{tabular}{lcc}
\hline\hline
  & Quantity & Unit\\
\hline
Diameter\tablefootmark{a} & 3.6 & pc \\ 
Width\tablefootmark{a} & 1.0 & pc \\
Mean column density ($\left<N_{\rm H}\right>$)\tablefootmark{b} & 3.8\,$\times$\,$10^{20}$ & cm$^{-2}$\\ 
Mass ($M_{\rm h}$) & 108 &  M$_{\odot}$\\
Expansion velocity ($\Delta v$)\tablefootmark{c} & 13.0 & km/s\\
Mean nucleon density ($\left<n\right>$) & 120 & cm$^{-3}$\\
\hline
\end{tabular}
\tablefoot{
\tablefoottext{a}{From radial profiles in Fig.~\ref{fig:radialprofiles}.} 
\tablefoottext{b}{Estimated for the emission in the range -5\,$<$\,$v_{\rm LSR}$\,$<$\,1\,\kps.}
\tablefoottext{c}{From $pv$ radial profiles in Fig.~\ref{fig:lvEONbubble}.}
}
\end{table}

\subsection{The EON bubble revealed by H{\sc i}}

Table~\ref{table:EONprops} presents the main physical properties of the EON shell derived from our H{\sc i} observations.
We estimated the approximate geometric center of the bubble in the velocity channel where its apparent angular size is largest, $v_{\rm LSR}$\,$\approx$\,1\,\kps, as shown in the left panel of Fig.~\ref{fig:bubbleRGB}.
After pinning the bubble center, around $l,b$\,$=$\,[209\pdeg1,$-$19\pdeg625], we computed radial profiles from which we estimated the size of the cavity, roughly 0\pdeg5, which corresponds to roughly 3.6\,pc at 414\,pc, as indicated in the profiles shown in top panel of Fig.~\ref{fig:radialprofiles}.
From these profiles, we infer that the walls of the shell are $\sim$1\,pc in width, although they are not homogeneous, as is evident in the scalloped regions toward the bottom rim and the EON protrusion.
The dust optical depth ($\tau_{353}$) profiles in the same directions, presented in the bottom panel of Fig.~\ref{fig:radialprofiles}, show that the shell walls are not clearly distinguishable in the integrated column density signal.

\subsubsection{EON shell column density}

We computed the hydrogen column density of the bubble for each pixel using equation~13.17 in \cite{wilson2013}, which we rewrote as
\begin{equation}\label{eq_NH}
N({\rm H}\textsc{i})=1.8224\times 10^{18} {\rm cm}^{-2} \sum^{[i_{0},i_{1}]}_{i} (T_{\rm b})_{i}\Delta v.
\end{equation}
$(T_{\rm b})_{i}$ is the \postlang{21 cm} line brightness temperature in Kelvin for the $i$th velocity channel.
$\Delta v$ is the velocity channel width in \kps.
The limits of the sum, $i_{0}$ and $i_{1}$, correspond to the minimum and maximum line-of-sight (LOS) velocities selected for the calculation.

We studied the column density in the front hemisphere of the EON shell by considering the emission in the range $-10$\,$<$\,$v_{\rm LSR}$\,$<$\,$1$\,\kps.
The lower velocity limit was set to the channel {where} the H{\sc i} brightness temperature {within} the bubble {fell} below detection levels.
The upper velocity limit corresponds to the emission where the bubble reaches its maximum extent.
This selection avoids confusion with the emission from the material in the OMC, which is prevalent at $v_{\rm LSR}$\,$>$\,$1$\,\kps, as illustrated in Fig.~\ref{fig:bubbleRGB}.

Equation~\eqref{eq_NH} assumes that the \postlang{21 cm} line emission is optically thin and there is no background emission.
These assumptions are justified for the bulk of the EON shell, but do not apply toward the Huygens region, where H{\sc i} absorption is observed against the radio continuum within a few arcminutes of the Trapezium stars \citep{vanderWerf2013}.
The $N({\rm H}\textsc{i})$ and mass estimates derived with Eq.~\eqref{eq_NH} might be contaminated by broad emission components produced by warm neutral medium (WNM) not directly associated with the bubble \citep[$T$\,$\sim$\,5000\,K][]{heiles2003,marchal2021}.
The separation of these components is not straightforward, but because the WNM has a relatively low density compared to the cold neutral medium (CNM) in the shell walls, the WNM is not expected to contain a significant amount of material, and its separation is not critical for the results of this paper. 
In any case, their subtraction would yield even lower $N({\rm H}\textsc{i})$ values, thus increasing the difference with the [C{\sc ii}]-based column densities.

Equation~\eqref{eq_NH} yields a mean H{\sc i} column density of $3.8$\,$\times$\,$10^{20}$\,cm$^{-2}$ for the 0\pdeg5-diameter region around the presumed bubble center and in the indicated velocity range.
Propagation of the measurement uncertainties in Eq.~\eqref{eq_NH} yields errors below the 1\,\% level, indicating that the variance in this value is dominated by fluctuations across the region and not by the observational uncertainty.
However, we note that H{\sc i} opacity effects can be considerable in regions where $T_{\rm b}$\,$\gtrsim$\,50\,K.
A full characterization of these effects requires constraints on the H{\sc i} excitation temperature and optical depth, which are derived from \postlang{21 cm} line observations in emission and in absorption against the radio continuum \citep[e.g.,][]{heiles2003,murray2018}.
The latter are limited to lines of sight with strong radio continuum sources and are not available for the VLA-D and FAST data combination.

The systematic study of the \postlang{21 cm} line optical depth across the Galactic plane presented in \cite{wang2020hi} indicated that the H{\sc i} column density is $\sim$40\% higher than that derived with the optically thin assumption behind Eq.~\eqref{eq_NH}. 
Studies of the H{\sc i} absorption toward the Perseus molecular cloud \citep[at around 300\,pc from the Sun;][]{bally2008a} \postlang{showed} that the correction for optical depth increases column density estimates by $\sim$10\% \postlang{\citep{lee2015}}.
Studies of synthetic \postlang{21 cm} line observations from multiphase numerical simulations indicate that the optically thin assumption can underestimate $N_{\rm H}$ by up to a factor of two \citep{seifried2022}.
Therefore, a conservative estimate suggests that line opacity effects might result in an underestimation of $N({\rm H}\textsc{i})$ by as much as a factor of two.

\subsubsection{EON shell mass}

\begin{figure}[h!]
\centering{
\includegraphics[width=0.48\textwidth,angle=0,origin=c]{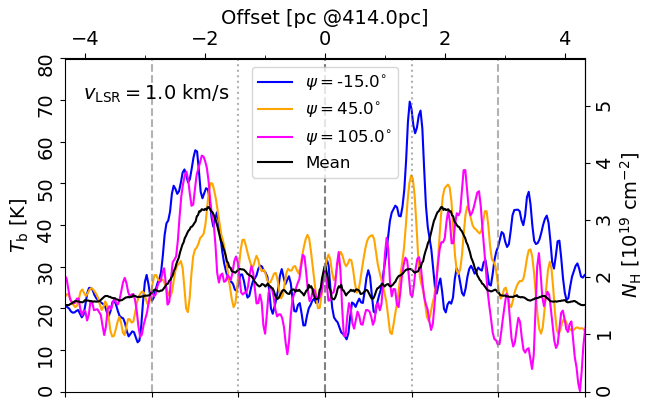}
\includegraphics[width=0.48\textwidth,angle=0,origin=c]{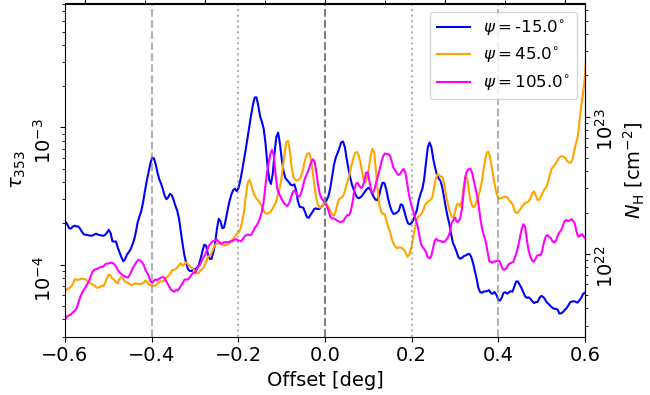}
}
\caption{Profiles of H{\sc i} emission radial profiles for $v_{\rm LSR}$\,$=$\,1\,\kps, roughly the LOS velocity showing the maximum bubble extension, and dust optical depth ($\tau_{353}$).
The three colored curves correspond to the directions indicated in the left panel of Fig.~\ref{fig:dusttauandT}.}\label{fig:radialprofiles}
\end{figure}

{We calculated the hemisphere mass by assuming that the column density estimated with Eq.~\eqref{eq_NH} is representative of the shell.
Multiplication of the hemisphere area by the mean surface density inferred from the H{\sc i} emission} results in 
\begin{equation}\label{eq:mass0}
M=2\pi r^{2}\left<N({\rm H}\textsc{i})\right>\mu\,m_{\rm H},
\end{equation}
where $r$ is the effective radius of the bubble, $\mu$ is the mean mass per hydrogen nucleus, and $m_{\rm H}$ is the hydrogen atom mass.
In a primarily atomic medium, $\mu$\,=1.42, which accounts for helium and the small contribution from metals for solar abundances \citep{asplund2021}.
In a partly molecular medium, values of $\mu$ are expected to be higher.  
When we use $r$\,$=$\,1.8\,pc, as inferred from Fig.~\ref{fig:radialprofiles}, Eq.~\eqref{eq:mass0} yields a mass of $\sim$108\,M$_{\odot}$ for the front side of the shell.
This mass estimate is significantly lower than that obtained from [C{\sc ii}] observations in \citetalias{pabst2020}, around 1,100\,M$_{\odot}$, as we discuss further in Sec.~\ref{sec:discussion}.

\subsubsection{EON shell expansion}

\begin{figure}[h!]
\centering{
\includegraphics[width=0.49\textwidth,angle=0,origin=c]{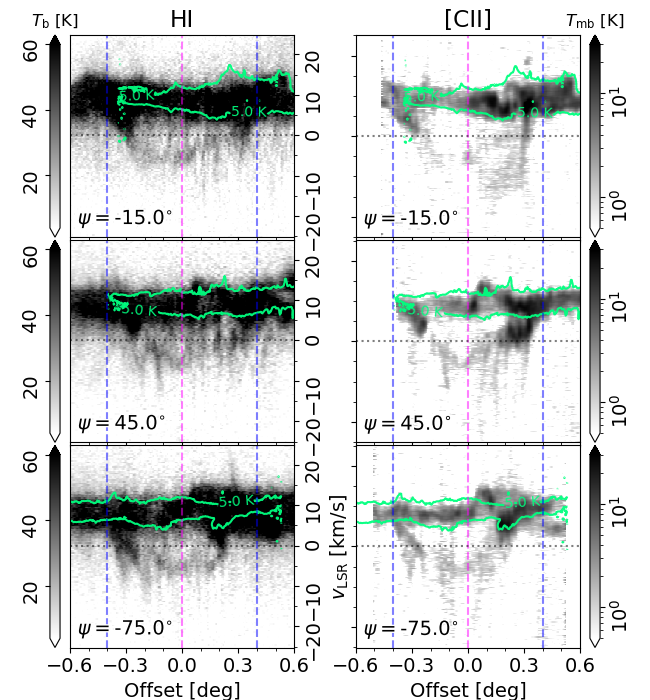}
}
\caption{H{\sc i} and [C{\sc ii}] emission across the radial profiles indicated in the left panel of Fig.~\ref{fig:bubbleRGB}.
The green contours show the 5-K level in the $^{12}$CO emission, corresponding to the OMC bulk.
}\label{fig:lvEONbubble}
\end{figure}

Figure~\ref{fig:lvEONbubble} presents position-velocity ($pv$) cuts through the H{\sc i} spectral cubes for the directions indicated in the left panel of Figure~\ref{fig:bubbleRGB}.
The H{\sc i} emission in the velocity channels corresponding to the bubble front, $v_{\rm LSR}$\,<\,1\,\kps, clearly shows the characteristic pattern of expanding shells, that is, a double component toward the center that progressively converges toward a single component toward the bubble rim \citep[e.g.,][]{heiles1979,mcclure-griffiths2002}.
The expansion pattern is very similar to that revealed by [C{\sc ii}] emission, {shown in the right column of Fig.~\ref{fig:lvEONbubble}}.
As indicated in \citetalias{pabst2020}, {the observer's side of the shell} shows no significant $^{12}$CO emission within the detection levels of the \cite{kong2018} observations, 0.86\,K.

{Figure~\ref{fig:lvEONbubble} shows that the H{\sc i} emission does not display the symmetric pattern expected for an idealized expanding bubble.
The approaching hemisphere shows the characteristic arc-like expansion pattern in the $pv$ cuts, but the receding hemisphere blends into the bulk of the OMC.}
This observation is consistent with the scenario of a blister-like bubble expanding from the molecular cloud \citep{odell2001,pabst2019}.
{As in these references, we used the velocity difference between the front-side emission and the bulk of the OMC to estimate an expansion velocity of $\sim$13\,\kps}.

A closer inspection of the H{\sc i} observations reveals what is potentially a second expanding shell within the EON region.
This feature is suggested by the anisotropy in the H{\sc i} emission at $v_{\rm LSR}$\,$\approx$\,$-1.8$\,\kps, shown in the middle panel of Fig.~\ref{fig:bubbleRGB}.
H{\sc i} emission at lower $v_{\rm LSR}$ progressively covers the semicircular cavity toward the lower portion of the EON, as also illustrated in Fig.~\ref{fig:bubbleRGB}.
This portion of the EON corresponds to one of the regions with excess diffuse emission in the X-ray 0.3 to 1-keV band reported in \cite{gudel2008}.

\begin{figure}[h!]
\centering{
\includegraphics[width=0.49\textwidth,angle=0,origin=c]{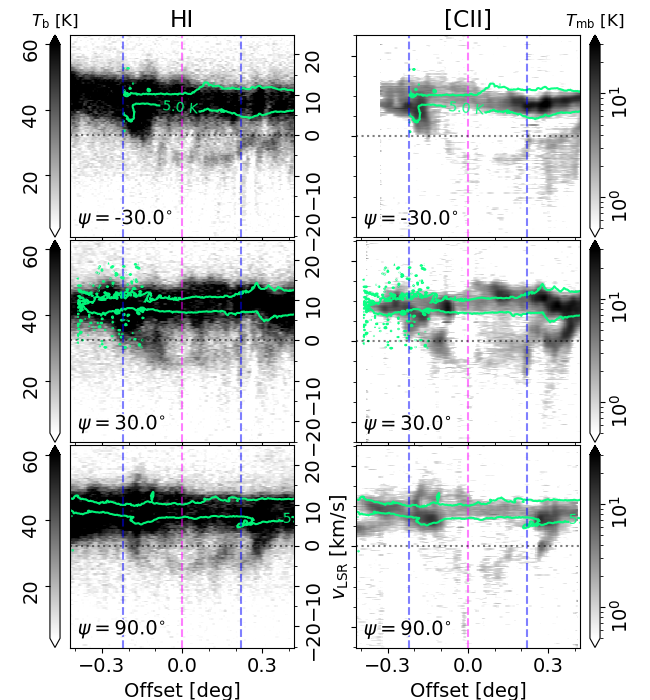}
}
\caption{Same as Fig.~\ref{fig:lvEONbubble}, but for the second EON bubble along the profiles shown in the middle panel of Fig.~\ref{fig:bubbleRGB}.   
}\label{fig:lvSECbubble}
\end{figure}

The H{\sc i} emission profiles toward the apparent secondary bubble, presented in Fig.~\ref{fig:lvSECbubble}, show two clear emission features between $-5$\,$<$\,$v_{\rm LSR}$\,$<$\,2\,\kps\ at $\pm$0\pdeg1 from the presumed center of the cavity.
These emission features might correspond to the walls of an expanding shell within the EON domain, whose limits are evident in the $pv$ profiles on either side of the secondary bubble.
The expansion velocity of the secondary bubble is approximately 10\,\kps, and, at the nominal distance of 414\,pc, it spans roughly 1.4\,pc in width.
The secondary bubble is not conspicuous within the detection limits of the [C{\sc ii}] emission observations, although its contours may be {apparent} in the profile shown in the middle panel of Fig~\ref{fig:lvSECbubble}.

\subsection{The EON protrusion}

The right panel of Fig.~\ref{fig:bubbleRGB} shows the elongated feature that we refer to as the EON protrusion.
This structure is much longer than what can be inferred from the [C{\sc ii}] emission, as illustrated by the bottom panels of Fig.~\ref{fig:multivpanels}.
At the standard distance to the Orion nebula, the protrusion extends around 4\,pc from the EON shell.
Its average column density, calculated using Eq.~\eqref{eq_NH}, is $\sim$9.9\,$\times$\,$10^{20}$\,cm$^{-2}$, and its peak column density is 1.6\,$\times$\,$10^{21}$\,cm$^{-2}$.
Its mass is around 80\,M$_{\odot}$, {as estimated from the} H{\sc i} emission {integrated} in the range $-15$\,$<$\,$v_{\rm LSR}$\,$<$\,7.5\,\kps.

\begin{figure}
\centering{
\includegraphics[width=0.49\textwidth,angle=0,origin=c]{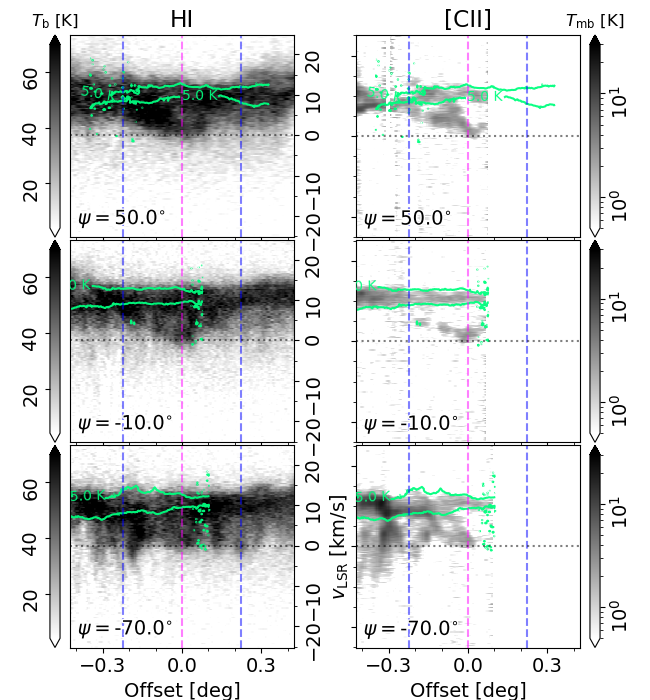}
}
\caption{Same as Fig.~\ref{fig:lvEONbubble}, but for the EON protrusion along the profiles shown in the right panel of Fig.~\ref{fig:bubbleRGB}.
The [C{\sc ii}] emission profiles are truncated by the coverage limits of the \citetalias{pabst2020} observations.
}\label{fig:lvEONchimney}
\end{figure}

The emission profiles across the EON protrusion, shown in the top and middle panels in Fig.~\ref{fig:lvEONchimney}, reveal an expanding-bubble-like pattern toward the center of the region in the right panel of Figure~\ref{fig:bubbleRGB}.
This apparently expanding region spans between $v_{\rm LSR}$\,$\approx$\,0\,\kps\ and the bulk of the CO emission line-of-sight velocity, $v_{\rm LSR}$\,$\approx$\,7.5\,\kps.
The location of this presumed bubble does not coincide with the position of any high-mass star or agglomeration of young stellar objects, {as discussed further in Sec.~\ref {sec:discussion:EONprotrusion}}.

The emission profiles along the main axis of the EON protrusion, shown in the bottom panel of Fig.~\ref{fig:lvEONchimney}, show a significant amount of emission throughout the structure in the range 0\,$\lesssim$\,$v_{\rm LSR}$\,$\lesssim$\,7.5\,\kps.
The plausible bubble-like patterns suggested by the [C{\sc ii}] emission in that profile are less evident in the H{\sc i}, which extends beyond the limits of the \citetalias{pabst2020} [C{\sc ii}] maps.
Toward the base of the EON protrusion, on the left side of the $pv$ cuts in the bottom panel of Fig.~\ref{fig:lvEONchimney}, the H{\sc i} and [C{\sc ii}] emission is present for $v_{\rm LSR}$\,$\lesssim$\,0\,\kps, but it does not appear to sample exactly the same structures. 

\section{Discussion}\label{sec:discussion}

The combination of the FAST and VLA reveals three observations that differ from the current understanding of the EON.
First, the mass derived from H{\sc i} emission is significantly lower than the values estimated from the [C{\sc ii}] line emission.
Second, the EON may not have been produced by a single source, the winds and the ionizing radiation from $\theta^{1}$\,Ori\,C, but at least one additional feedback event may have shaped the region. 
Finally, the H{\sc i} maps show an elongated structure extending from the bubble, suggesting that material from the {primary shell is potentially being} disrupted.

\subsection{Mass of the EON shell}

Extinction toward $\theta^{1}$\,Ori\,C was established to be about $A_{\rm V}$\,$=$\,1.6 in \cite{johnson1967}.
Subsequent studies showed that the reddening curve for the region was flatter than was typical in the interstellar medium (ISM), complicating the determination of column density and elemental abundances \citep[e.g.,][]{cardelli1988,blagrave2007}.
Observations of Lyman $\alpha$ (Ly$\alpha$) absorption indicated H{\sc i} column densities around 10$^{21}$\,cm$^{-2}$ in {Orion's veil}, the layer of material in front of the Trapezium cluster \citep{savage1972,savage1977}.
Measurements of He{\sc i} and Ca{\sc i} optical absorption lines toward Trapezium stars indicate values higher by at least a factor of four \citep{odell1993,shuping1997}.
However, the conditions of {Orion's veil} cannot be directly extrapolated to the larger and more diffuse EON.

Using observations from ESA's X-ray space telescope XMM-Newton, \cite{gudel2008} derived an absorbing hydrogen column density of $N({\rm H})$\,$\approx$\,4\,$\times$\,10$^{20}$\,cm$^{-2}$ toward the northern EON and $N({\rm H})$\,$\approx$\,10$^{20}$\,cm$^{-2}$ toward the southern X-ray emitting region.
These values agree with those derived from our combination of FAST and VLA H{\sc i} observations.
However, they differ from those obtained from [C{\sc ii}] line emission.

\citetalias{pabst2020} derived a [C{\sc ii}] column density $N(\rm{C}\textsc{ii}{})$\,$=$\,3\,$\times$\,10$^{17}$\,cm$^{-2}$.
The corresponding H column density can be inferred from the carbon nucleon budget in the region,
\begin{equation}\label{eq:NHfromC}
  N({\rm H}) = \frac{N({\rm C})}{X({\rm C})} = \frac{N({\rm C}\textsc{i})+N({\rm C}\textsc{ii})+N({\rm C}\textsc{iii})+N({\rm CO})+N(\rm{CX})}{X({\rm C}},
\end{equation}
where $X({\rm C})$ represents the elemental abundance ratio of C relative to H, and the terms in the sum correspond to the column densities for neutral atomic carbon (C{\sc i}), singly and double ionized carbon (C{\sc ii} and C{\sc iii}), and other carbon-bearing species (CX).
When we assume that all of the carbon along the line of sight is singly ionized and that there is no significant amount of carbon in CO and other carbon molecules, Eq.~\eqref{eq:NHfromC} reduces to
\begin{equation}\label{eq:NHtoNC}
  N({\rm H}) \approx N({\rm C}\textsc{ii})/X({\rm C}).
\end{equation}
\citetalias{pabst2020} assumed $X({\rm C})$\,$=$\,1.6\,$\times$\,$10^{-4}$ from \cite{sofia2004}, which yields a total H nuclei column density N({\rm H})\,$\approx$\,1.87\,$\times$\,$10^{21}$\,cm$^{-2}$. This is higher by about a factor of five than that obtained from the H{\sc i} emission for the front of the shell.

The analysis of UV observations toward the Orion nebula presented in \cite{rubin1993} indicates values of $X({\rm C})$\,$=$\,2.8\,$\times$\,$10^{-4}$.
This value is closer to the solar abundance \citep[2.9\,$\times$\,$10^{-4}$,][]{asplund2021} than that used in \citetalias{pabst2020}.
These alternative $X({\rm C})$ values reduce the difference between the [C{\sc ii}]-derived $N(\rm{H})$ and the \postlang{21 cm} $N(\rm{H}\textsc{i})$ values to a factor of around two, but they do not fully reconcile the two measurements.

{\citetalias{pabst2020} estimated the [C{\sc ii}] optical depth and column density using the peak temperature of the [$^{12}$C{\sc ii}] line and the [$^{13}$C{\sc ii}]$F$\,$=$\,2\,$\rightarrow$\,1 hyperfine line \citep{boreiko1996,ossenkopf2013}}. 
The [$^{12}$C{\sc ii}] line traces material at $v_{\rm LSR}$\,$\approx$\,8.8 and 4.0\,\kps, as shown in their figure~3, while the [$^{13}$C{\sc ii}]$F$\,$=$\,2\,$\rightarrow$\,1 hyperfine line is only detected at $v_{\rm LSR}$\,$\approx$\,8.2\,\kps. 
This means that their estimates include dense material from the OMC, which may not be representative of the front shell revealed by the H{\sc i} emission at $v_{\rm LSR}$\,$<$\,1\,\kps.
By assuming that the limb-brightened parts of the shell, at $v_{\rm LSR}$\,$>$\,1\,\kps, applied to the frontal hemisphere, the analysis in \citetalias{pabst2020} may be overestimating the shell mass.

\subsection{The H$_{2}$ possibility}

\postlang{Although the arguments above suggest that the analysis of the [CII] observations by \citetalias{pabst2020} may overestimate $N({\rm H})$ in the front shell hemisphere, part of the discrepancy could arise from molecular material in the shell. 
Assuming a homogeneous hemispherical shell with radius 2.7\,pc and width 0.3\,pc, the 1100\,\msun\ mass estimated by \citetalias{pabst2020} implies a mean nucleon density of $\sim$3.2,cm$^{-3}$, corresponding to a column density of $\sim$3\,$\times$\,$10^{21}$\,cm$^{-2}$ across the shell. 
This value, or the column density along the longest straight-line path that remains entirely within the shell, 2.6\,$\times$\,$10^{22}$\,cm$^{-2}$, is below the fluctuations in the dust-derived $N({\rm H})$ profiles shown in the bottom panel of Fig.~\ref{fig:radialprofiles}. 
This implies that the integrated dust opacity does not conclusively exclude such column densities in the shell, leaving open the possibility of sufficient shielding for a significant fraction of the hydrogen to remain in molecular form.}

The ratio of H and C nucleons does not distinguish between ionized (H{\sc ii}), atomic, and molecular hydrogen (H$_2$).
Thus, the difference between the column densities inferred from H{\sc i} and [C{\sc ii}] line emission suggests that at least part of the H nuclei in the shell are in the form of H$_2$.
The hydrogen nucleon column density can be decomposed as
\begin{equation}\label{eq:NHfromH}
N({\rm H})=N({\rm H}\textsc{i})+N({\rm H}\textsc{ii})+2N({\rm H}_{2}),
\end{equation}
where each term in the sum corresponds to the column densities for hydrogen in the atomic, ionized, and molecular phases.
According to estimates based on H$\alpha$ emission in \citetalias{pabst2020}, $N({\rm H}\textsc{ii})$ contributes to less than 2\% of N({\rm H}), so we considered it negligible.
Combining Eq.~\eqref{eq:NHfromH} with the equivalent estimate from the carbon budget using Eq.~\eqref{eq:NHtoNC}, we obtained 
\begin{equation}\label{eq:NH2}
N({\rm H}_{2})=\frac{1}{2}\left(\frac{N({\rm C}\textsc{ii})}{X({\rm C})}-N({\rm H})\right).
\end{equation}
When we take the \citetalias{pabst2020} estimates as representative for the whole shell along with our $N({\rm H{\textsc{i}}})$, Eq.~\eqref{eq:NH2} yields a molecular hydrogen column density $N({\rm H}_{2})$\,$\approx$\,3\,$\times$\,10$^{20}$\,cm$^{-2}$, which corresponds to a molecular-to-atomic hydrogen fraction, $N({\rm H_{2}})/N(\rm H)$\,$\approx$\,0.8.
This value is relatively high compared to the Milky Way average of 0.2 to 0.3 \citep[inferred from Galactic H{\sc i} and CO masses;][]{kalberla2009,heyer2015}, as expected for the relatively high-density region in front of the OMC.

Observations of H$_{2}$ UV absorption toward the OMC are almost exclusively concentrated on the Trapezium stars \citep[see][and references therein]{bellomi2020}.
Measurements with the ultraviolet scanning spectrometer of the {\tt Copernicus} satellite telescope identified $N({\rm H}_{2})$\,$<$\,3.5\,$\times$\,$10^{17}$\,cm$^{-2}$ in Orion's veil \citep{savage1977}, implying a relatively low fraction of molecular gas, $N({\rm H}_{2})/N({\rm H})$\,$\lesssim$\,10$^{-4}$.
However, these conditions cannot be directly extrapolated to the EON shell, where the larger distance to the Trapezium stars guarantees a lower UV field and potentially a population of smaller dust grains that favors H$_{2}$ formation \citep[e.g.,][]{cazaux2004} and maintains the warm molecular gas component expected in low-extinction regions where photoelectric heating remains effective \citep[e.g.,][]{glover2010,wolfire2010}.

\citetalias{pabst2020} estimated a mean volumetric nucleon density, $n$\,$\sim$\,$10^{3}$ cm$^{-3}$ for the EON shell.
At this density, it is plausible that the material is sufficiently shielded to become and remain molecular, despite the absence of CO emission toward the observer's side of the shell and at the bubble rim.
However, this $n$ value was obtained assuming the $X({\rm C})$ value from \citep{sofia2004}, which accounts for dust depletion and may not be representative of the entire EON region.
The $X({\rm C})$ values from \cite{rubin1993} for the Orion nebula or solar abundances would yield lower $n$ by a factor of $\sim$2.

\citetalias{pabst2020} assumed a shell thickness $\sim$0.3\,pc, in contrast with the 1\,pc we inferred from the H{\sc i} radial profiles in Fig.~\ref{fig:radialprofiles}.
The former value was computed from the inspection of the emission in the shell rim in the $v_{\rm LSR}$ range between 0 and 5\,\kps\ in 0.2-\kps wide channels.
The latter value was computed from the azimuthal average of the H{\sc i} emission at the $v_{\rm LSR}$ where the bubble showed its maximum extension and minimum blending with the surrounding material. 
When we divide the [C{\sc ii}]- and H{\sc i}-derived column densities by the 1 pc thickness, we obtain $n$\,$\sim$\,330 and 120\,cm$^{-3}$, respectively.
These values suggest that the shell is composed of a more translucent medium than implied by the \citetalias{pabst2020} estimations.
However, their computation neglected the limb-brightening effect evident in the emission maps of Fig.~\ref{fig:multivpanels}, which likely {results} in a projected size larger than the actual shell thickness.
Thus, the above values are lower limits of the shell column density and still leave room for a some amount of H$_{2}$.

{The observations in \cite{dame2001} and \cite{kong2018} indicated that $^{12}$CO$(1\,$\,$\rightarrow$\,$0)$ line emission is largely absent in the velocity range of the EON front shell hemisphere, as illustrated in Fig.~\ref{fig:lvEONbubble}.
Observations of $^{12}$CO$(2\,$\,$\rightarrow$\,$1)$ at 11\arcsec\ resolution presented in \cite{goicoechea2020} indicated the presence of several CO globules blueshifted in velocity with respect to the OMC and embedded in the [C{\sc ii}]-bright shell, but they represent less that 3\% of its mass.
Thus, at the moment, evidence of a significant amount of molecular gas to reconcile the shell masses derived from H{\sc i} and [C{\sc ii}] line emission is not available.
}

\subsection{EON shell models}

Based on the shell size and expansion velocities reported in Table~\ref{table:EONprops}, the EON shell dynamical timescale is around 0.25\,Myr.
This is within the expected range for bubbles blown by main-sequence O stars, which is between 0.1 and 1\,Myr according to standard \cite{weaver1977} models, assuming an adiabatic interior and negligible external pressure.

When we consider the EON front hemisphere {as one half of a} homogeneous spherical shell, we estimate its kinetic energy as
\begin{equation}
E_{\rm k}=\frac{1}{2}\left(2M_{\rm h}\right)\left(\frac{\Delta v}{2}\right)^{2},
\end{equation}
{where $M_{\rm h}$ and $\Delta v$ are the half-shell mass and bubble expansion velocities presented in Table~\ref{table:EONprops}.
The resulting value, $\sim$10$^{47}$\,erg, is} within the expected range for a wind-blown bubble in the \cite{weaver1977} models, which yield $E_{\rm k}$\,$\approx$\,0.6\,$\times$\,$10^{47}$ to 3.4\,$\times$\,$10^{47}$\,erg for initial cloud-to-intercloud mass ratios 1 and 0.01 \citep{pittard2022}.

{Assuming axial symmetry, the shell momentum is} 
\begin{equation}
p=2M_{\rm h}\left(\frac{\Delta v}{2}\right),
\end{equation}
{which yields} $\sim$7\,$\times$\,10$^{3}\,$\,M$_{\odot}$\,\kps.
This value is slightly higher than the range of 1.9 to 5.8\,$\times$\,10$^{3}$\,M$_{\odot}$\,\kps\ obtained with the \cite{weaver1977} models.
A higher implied mass, as can be obtained by folding in a factor of 2 for H{\sc i} opacity or considering the [C{\sc ii}]-derived values, would increase this gap.
{We note, however, that the \cite{weaver1977} models are axisymmetric and do not account for the fact that the EON shell is expanding against the OMC.}

We further compared our observations with semi-analytical models using a grid of models in the one-dimensional (1D) bubble-evolution code {\tt TRINITY}, which computes the evolution of feedback-driven bubbles through energy- and momentum-driven phases while self-consistently coupling all relevant pressure sources. 
{\tt TRINITY} extends the 1D cloud evolution framework of WARPFIELD \citep{rahner2019}, solving the coupled equations for shell radius, hot bubble energy, and swept-up mass with time-dependent stellar wind and radiation inputs from stellar population synthesis.

For the {\tt TRINITY} modeling of the EON shell, we assumed a cloud with a homogeneous density profile, solar metallicity, and mean initial densities $n_{\rm core}$\,$=$\,$10^{2}$, 5\,$\times$\,$10^{2}$, and 10$^{3}$\,cm$^{-3}$. 
To match the properties of $\theta^1$\,Ori\,C ($M_\star$\,$=$\,34\,$\pm$\,5\,\msun), we constructed a grid of models spanning cloud masses ($M_{\rm cl}$) between $10^3$ and 5\,$\times$\,$10^3$\,\msun\ and star-formation efficiencies ($\epsilon$) between 0.68 and 3.4\%, selecting the eight tracks that yielded stellar masses in the $\theta^1$\,Ori\,C range.
{\tt TRINITY} is designed to simulate cluster-driven feedback, so we derived time-dependent feedback parameters, that is, bolometric luminosity, mechanical luminosity, and ionizing photon rate, from {\tt Starburst99} \citep{leitherer1999} using Geneva nonrotating tracks \citep{ekstrom2012} for a $10^6$-\msun\ cluster, and we then scaled linearly to $34$\,\msun. 
This approach captures the temporal evolution of feedback but may overestimate the integrated energy input, since massive clusters include stars across the initial mass function (IMF) rather than a single dominant source.

\begin{figure}[h!]
\centering{
\includegraphics[width=0.48\textwidth,angle=0,origin=c]{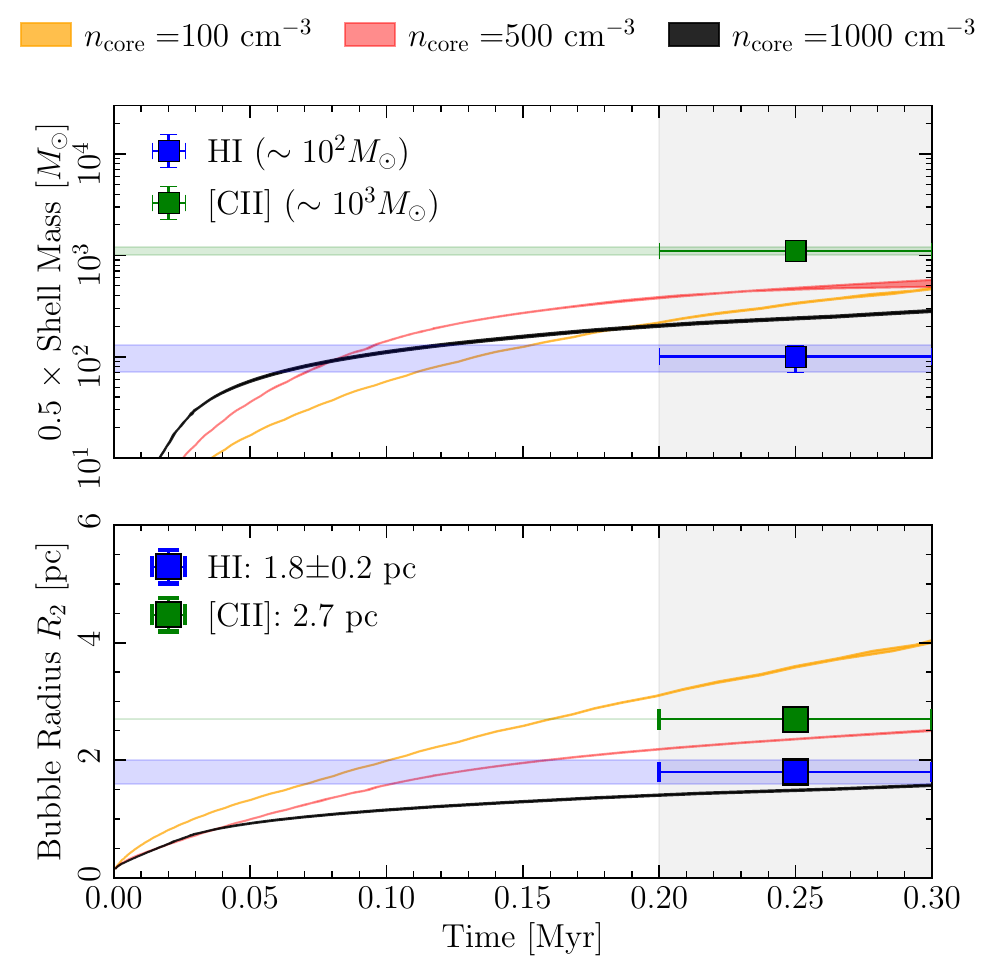}
}
\caption{Half-shell mass (top) and radius (bottom) evolution from {\tt TRINITY} models of the EON for initial densities $n_{\rm core}$\,$=$\,$10^{2}$, 5\,$\times$\,$10^{2}$, and 10$^{3}$\,cm$^{-3}$ (orange, red, and gray curves). 
The horizontal bands indicate observational constraints: the H{\sc i}-derived mass ($\sim 10^2$\,\msun; {blue}) and [C{\sc ii}]-derived mass ($\sim$$10^3$\,\msun; {green}) in the top panel.
}
\label{fig:trinitymodels}
\end{figure}

Figure~\ref{fig:trinitymodels} shows the resulting {shell hemisphere} mass and radius evolution for the selected models.
They are comparable with the \citep{castor1975} model employed in \citetalias{pabst2020} to infer a shell dynamic time of 0.24\,$\pm$\,0.05\,Myr and an initial density (1.7\,$\pm$\,0.7)\,$\times$\,$10^{2}$\,cm$^{-3}$ from their inferred shell radius, expansion velocity, mass, plasma temperature, and density.
In our application, we used {\tt TRINITY} to forward-model the shell from generic initial conditions. 

The shell radius evolution obtained with {\tt TRINITY}, shown in the bottom panel of Fig.~\ref{fig:trinitymodels}, indicates that the size and dynamical timescales derived from the H{\sc i} observations are roughly consistent with the evolution of the model.
The shell mass evolution obtained with {\tt TRINITY}, shown in the top panel of Fig.~\ref{fig:trinitymodels}, indicates values that are {between the masses derived from the H{\sc i} and [C{\sc ii}] observations}.
{An increase in the initial density within the explored range does not change the shell mass enough to reconcile the [C{\sc ii}]-derived mass with the models.}

{The results of the TRINITY model suggest that the [C{\sc ii}] analysis in \citetalias{pabst2020} might overestimate the mass of the half shell, which is consistent with the [C{\sc ii}] line opacity complications we  discussed so far.
We acknowledge the model limitations in fully reproducing the EON shell, however. 
The feedback scaling in {\tt TRINITY} likely overestimates the energy input, and the actual shell radius might be smaller than the model predicts.
Moreover, TRINITY models the evolution of a bubble produced by a 34\,\msun\ ionizing mass, rather than a cluster. 
This treatment excludes the contributions from other members, such as $\theta^{2}$\,Ori\,A. 
Implementing this is a significant addition to the existing code, as it involves coupling TRINITY to a stochastic IMF sampling code, which is planned for future work but is not available at the moment.
Additionally, a quantitative prediction of the expected H$_{2}$ mass would require coupling {\tt TRINITY} to PDR models. This capability is not yet available.
Finally, another limitation stems from our observational understanding of the EON shell, which may be more complex than a perfectly axisymmetric spherical wind-blown bubble.
}

\subsection{EON shell substructure}

The anisotropy in the H{\sc i} emission distribution shown in the middle panel of Fig.~\ref{fig:bubbleRGB} and the features in the $pv$ profiles in Fig.~\ref{fig:lvSECbubble} suggest that there is a secondary expanding structure within the EON shell.
The soft X-ray emission produced by the million-degree plasma pervading the EON cavity identified in \cite{gudel2008} is distributed in two main hubs, like the head and body of a snowman, in the western portion of the region in equatorial coordinates.
The division between these two hubs was attributed to an extinction feature in the front bubble hemisphere. 
However, the H{\sc i} emission toward the observer's side of the region shows no structure that could produce this division. 
It shows instead that the southern hub of diffuse X-ray emission coincides with an expanding cavity.

The main and secondary EON bubbles might have been produced by two consecutive feedback events. 
First, the main EON bubble is blown by the winds from $\theta^{1}$\,Ori\,C.
Second, another high-mass star leaving the ONC produces feedback, even shaping the second bubble.
\cite{kim2019} reported proper motions of around 1.4\,$\times$\,$10^{-3}$ arcseconds per year for stars in the ONC, around 3\,pc\,Myr$^{-1}$.
Thus, it is plausible that a star traveled the $\sim$2\,pc from the cluster to the center of the secondary bubble within the 5 to 7\,Myr main-sequence lifetime of an O7 star such as $\theta^{1}$\,Ori\,C.
The 1.4-pc diameter of the EON secondary bubble is smaller than the expected final cavity size for a supernova (SN) in standard ISM conditions \citep[e.g.,]{martizzi2015}.
However, the presumed precursor of this secondary bubble is exploding inside a preexisting cavity, implying a much larger diameter. 
The only way to reconcile the secondary EON bubble with the SN scenario is to assume it is a very young supernova, younger than 300 years old. An event like this, however, so close to the Sun, could not have gone unnoticed.

\begin{figure*}[h!]
\centering{
\includegraphics[width=0.95\textwidth,angle=0,origin=c]{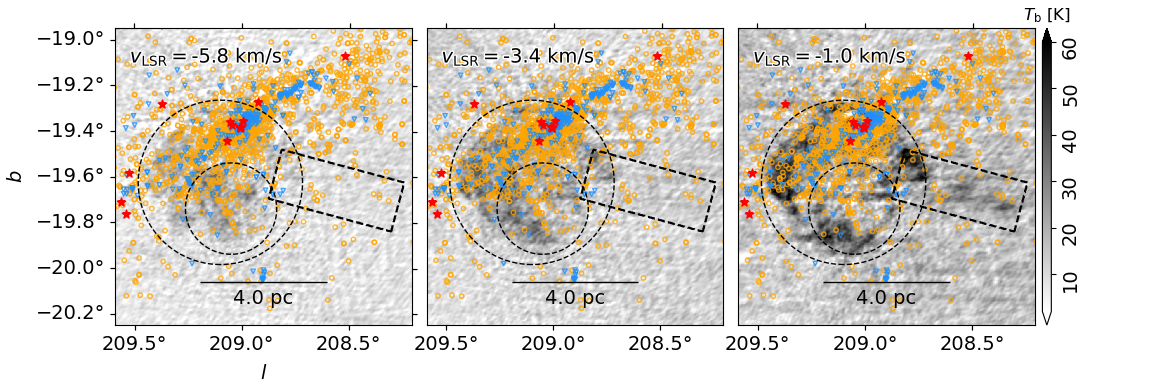}
}
\caption{H{\sc i} emission for velocity channels showing the secondary EON bubble.
The red stars indicate the position of OB stars in the \cite{pantaleoni-gonzalez2021} catalog.
The orange circles and blue triangles show the positions of the class 0/I and class II objects in the \cite{roquette2025} 
YSO catalog.
The dashed black circles correspond to the estimated maximum extension of the EON shell and the presumed secondary bubble. 
The dashed rectangle indicates the location of the EON protrusion.
}\label{fig:multivSECbubble}
\end{figure*}

{An alternative explanation is feedback from high-mass stars on the main sequence.}
Fig.~\ref{fig:multivSECbubble} presents the H{\sc i} emission for a set of velocity channels showing the EON secondary bubble along with the positions of O- and B-type stars whose winds or radiation could push such a bubble.
We found no evident candidate for the progenitor of the secondary bubble among the stars in the \cite{pantaleoni-gonzalez2021} OB stars catalog, one of the most complete in terms of photometric and spectroscopic classification, or a concentration of YSOs in the \cite{roquette2025} catalog, as illustrated in Fig.~\ref{fig:multivSECbubble}.

The absence of a progenitor at the center of bubbles is not a fundamental requirement for the validation of a feedback-blown bubble. 
The presumed progenitor of the EON, $\theta^{1}$\,Ori\,C, is not at the center of the cavity, which in other observational evidence has been interpreted as an indication that another star, O9.5IVp-type {$\theta^{2}$\,Ori\,A}, is the dominant source outside of the Orion bar and as far as the southeast EON boundary \citep{odell2017}.
It is plausible that runaways and walkaway stars from the ONC have impacted their surrounding ISM \citep[see, for example,][]{fujii2022}.
However, because the velocity dispersions of stars \citep{kim2019} and the H{\sc i} gas are so similar, it is almost impossible to find a conclusive relation between the source and the H{\sc i} shell.



\subsection{The EON protrusion}\label{sec:discussion:EONprotrusion}

Figure~\ref{fig:multivpanels} shows a prominent structure protruding from the EON shell across 1\,$<$\,$v_{\rm LSR}$\,$<$\,5\,\kps.
\postreftwo{\cite{kavak2022protrusion}} identified a protrusion-like substructure in the [C{\sc ii}] observations toward the northwestern portion of the EON shell.
The location and orientation of this structure coincide with the base of the elongated feature revealed by H{\sc i} for $v_{\rm LSR}$\,$\approx$\,0\,km/s, shown in Fig.~\ref{fig:EONhiANDwise}, which reaches beyond the \postreftwo{extent} of the [C{\sc ii}] observations.

\cite{kavak2022protrusion} identified jet-like elongated structures in the near-infrared \postlang{\it Spitzer} and WISE  observations toward the [C{\sc ii}] protrusion. 
\postlang{Based on the morphology of these features and estimates of kinetic energy, the authors concluded that this portion of the EON shell corresponds to a region of the preexisting cloud locally perturbed by outflows from massive protostars, suggesting that protrusion results from mechanical and not radiative feedback.}
The elongated structure revealed by our H{\sc i} observations reinforces the hypothesis of a preexisting cloud in the location of the protrusion. 
However, it is unlikely that the elongation of this H{\sc i} structure is explained by the outflows from massive protostars. 

\cite{kavak2022protrusion} also suggested that the protrusion is a suitable place to break Orion’s veil owing to the photo-ablation from the shell walls.
The authors identified what appear to be two half-shells with expansion velocities of 6 and 12\,\kps\ at the base of the protrusion.
Our H{\sc i} observations show abundant emission throughout the protrusion in the velocity range in which these bubbles were identified.
This observation does not contradict the presence of two or more bubbles along the protrusion, but it suggests that additional significant gas dynamics are present in that structure.

Shells bounding expanding wind-blown bubbles can fragment due to the nonlinear growth of the Rayleigh-Taylor, thermal, and other instabilities, as illustrated by the scalloping of the EON shell shown in Fig.~\ref{fig:EONhiANDwise}.
The hot high-pressure gas can rupture the inhomogeneous shell and rapidly flow into the undisturbed environment, as shown, for example, in numerical experiments in \cite{geen2022} and \cite{geen2023}.
As the hot gas propagates through holes in the main shell, it ablates material at the interfaces between holes and fragments in the main shell.
The expansion of the hot gas in front of the main shell also sweeps up ambient material into a blister shell, which is consistent with the expanding-shell structures identified in \postreftwo{\cite{kavak2022dents}}.
However, the size of these blisters at the surfaces is expected to be smaller than the size of the bubble \citep[see, for example,][]{pittard2013}, which is not the case for the $\sim$4-pc diameter EON shell and the $\sim$4-pc long EON protrusion revealed by the H{\sc i}.

Globally, the EON shell retains its almost circular shape, so it is likely that the potential blowout at the base of the EON protrusion did not cause a catastrophic depressurization. 
Thus, it is plausible that the puncture has been self-sealed by the cooling swept-up gas.
Still, the buildup of the EON protrusion via consecutive outflows from the leaky shell requires that these outflows always occur in the same direction. 
The lack of a clear velocity gradient along the main axis of the structure, as shown in the bottom left panel of Fig.~\ref{fig:lvEONchimney}, indicates that this structure does not evidently flow into or from the EON shell.

Orthogonal evidence for the nature of the EON protrusion comes from thermal dust emission toward the region, for which we used the dust opacity and temperature obtained from the combination of {\it Planck} and {\it Herschel} observations presented in \cite{lombardi2014}.
The dust opacity map, shown in the left panel of Fig.~\ref{fig:dusttauandT}, does not show a conspicuous presence of the EON shell walls, most likely due to the dust in its background.
The dust temperature map, presented in the right panel of Fig.~\ref{fig:dusttauandT}, reveals the higher dust temperature in the bubble walls, most likely heated by radiation from the ONC cluster.
This high average temperature along the line of sight does not extend in the direction of the protrusion, suggesting that the heating source of the EON shell does not affect the protrusion. 
Moreover, the average dust temperature across the protrusion is low, which either indicates that it is a preexisting density structure with dust at a temperature similar to the background, or that it is a pure H{\sc i} structure with little dust along it.



\section{Conclusions}\label{sec:conclusions}

We presented new arcminute-resolution H{\sc i} emission maps toward the EON region that were obtained by combining VLA and FAST observations.
These maps reveal the neutral atomic hydrogen counterpart to the expanding bubble previously suggested by soft X-ray and [C{\sc ii}] observations.
They also display additional structures that suggest that the bubble is not a single wind-blown cavity.

The H{\sc i} observations reveal for the first time the front hemisphere of the EON shell, for which we estimate a mass $\sim$100\,M$_{\odot}$.
This value is lower by about a factor of {ten} than that obtained for one half of the shell using [C{\sc ii}] emission.
Given the wide integration ranges and averaging of physical conditions necessary to produce this estimate from the [C{\sc ii}] line, this constitutes the most direct measurement of the mass displaced by the expansion of the EON bubble.

The discrepancy between the H{\sc i}- and [C{\sc ii}]-derived shell masses might also imply that a significant amount of the material in the EON shell is in the form of H$_{2}$.
Testing this possibility requires additional studies of UV absorption and infrared quadrupole emission that directly reveal the H$_{2}$ associated with the multiphase structure we studied using H{\sc i} and [C{\sc ii}].
Our observations revealed that the rim of the EON shell is an ideal laboratory for the study of the PDR and the H{\sc i}-to-H$_{2}$ transition.

The \postlang{21 cm} emission also reveals a potential secondary bubble within the main EON shell.
This presumed structure is evident in the anisotropy of the H{\sc i} emission on the observer's side of the shell.
Although a progenitor for this structure is not identified, its presence indicates that the EON shell is not exclusively the product of stellar winds from $\theta^{1}$\,Ori\,C star.

The H{\sc i} maps show an elongated structure protruding 4-pc from the EON shell. 
This protrusion with a mass $\sim$80-M$_{\odot}$ might be related to the outflow of hot gas from the EON.
However, its extension and kinematic structure suggest that it is not produced solely by stellar feedback in the region and might be part of a preexisting structure in the vicinity of the EON.

Our observations revealed previously uncharted features in the nearest wind-blown bubble and high-mass star-forming region.
They also demonstrated the potential of Galactic H{\sc i} studies, combining single-dish and interferometric observations, to reveal unexplored gas dynamics and provide a more complete picture of the interactions between star-forming regions and their surroundings. 
Even in a well-studied region such as Orion, H{\sc i} reveals something new in the heavens.


\section*{Data availability}
The combined H{\sc i} datacube is available at the CDS via anonymous ftp to
\href{ftp://cdsarc.u-strasbg.fr}{cdsarc.u-strasbg.fr} (130.79.128.5)
or via the
\href{http://cdsweb.u-strasbg.fr/cgi-bin/qcat?J/A+A/}{CDS catalogue}.

\begin{acknowledgements}

The ``Neutral Atomic Hydrogen in the solar neighborhood'' (NeAtHood) project is funded by the Austrian Science Fund (Fonds zur Förderung der wissenschaftlichen Forschung, FWF) through Grant DOI 10.55776/PAT6169824 (PI: J.~D.~Soler).
JDS thanks the following people for their encouragement and conversation: Francesca Bonanomi, Sergio Dzib, Bruce Elmegreen, Adam Ginsburg, Manuel G\"{u}del, Antoine Marchal, and Naomi McClure-Griffiths.
JDS thanks Marko Kr\v{c}o and Di Li for their assistance with the FAST CH{\sc i}NA data.  
We also thank the anonymous referee for carefully reading our manuscript and providing insightful comments and suggestions.
This research was carried out in part at the Jet Propulsion Laboratory, which is operated by the California Institute of Technology under a contract with the National Aeronautics and Space Administration (80NM0018D0004).
SCOG acknowledges financial support from the European Research Council via ERC Synergy Grant ``ECOGAL'' (project ID 855130) and from the German Excellence Strategy via the Heidelberg Cluster ``STRUCTURES'' (EXC 2181 - 390900948). 
DS acknowledges support of the Bonn-Cologne Graduate School, which is funded through the German Excellence Initiative as well as funding by the Deutsche Forschungsgemeinschaft (DFG) via the Collaborative Research Center SFB 1601 ``Habitats of Massive Stars Across Cosmic Time'' (subprojects B1 and B4).
{\it Software}: {\tt astropy} \citep{astropy2018}, {\tt magnetar} \citep{magnetar2020}.
\end{acknowledgements}

%
\bibliographystyle{aa}
\bibliography{aa59272-26corr.bbl}

\begin{appendix}





\section{VLA data preparation}\label{app:VLA}

The ODIN correlator configuration is designed to sample the H{\sc i} \postlang{21 cm} line in a 1-MHz-wide spectral sub-band.
To obtain the line-free portions of the spectrum necessary for continuum subtraction, we sampled the range between $\pm$50\,\kps\ in 0.4-\kps-wide (1.89-kHz) channels.
Four additional 1-MHz spectral sub-bands were used to sample four OH hyperfine structure lines at 1612, 1665, 1667, and 1720\,MHz at 0.4\,\kps\ resolution.
An additional 2-MHz-wide set of sub-bands was used to sample 13 H-$\alpha$ radio recombination lines (RRLs) at 10-\kps\ resolution.
Finally, ODIN also covered the L-band continuum emission in full polarization in seven broad sub-bands, which cover the radio-frequency interference (RFI)-free portions of the spectrum.
The analysis of these additional data products will be presented in a separate publication.

The ODIN data were flagged and calibrated using the VLA's scripted calibration pipeline.
The pipeline's goal is to obtain optimal calibration solutions for all SPWs in the ODIN observations.
However, the original pipeline was modified to optimize the H{\sc i} observations.
For example, the initial Hanning smoothing was disabled to preserve the original spectral resolution.
Final automatic flagging of the target observations was disabled to protect the line spectral window (SPW).


Data flagging was implemented as described in \citet{bihr2015}.
Strong RFI and poor antennas were manually flagged before calibration.
The VLA pipeline routine applied additional automated RFI flagging to the calibrators.
Additional manual RFI flagging was applied later during the image processing.


The flux, bandpass, and polarization calibrations were performed using 3C\,48 as a reference source. 
This object's spectrum shows a dip likely produced by H{\sc i} absorption \citep{murray2015}.
We circumvented this problem by applying polynomial interpolation to the bandpass. 
This solution mitigates the most evident artifacts from this feature in the calibrator, which may produce line emission appearing at exactly the calibrator’s absorption velocity, absorption/emission features in every source in the field, or identical spectral structure in targets with very different physics.
If still present, these effects are smaller than the signal produced by H{\sc i} in the EON shell and do not significantly affect the conclusions of our study. 


Once calibrated, the target H{\sc i} scans were separated from the remaining ODIN data.
The continuum emission was subtracted in the $uv$-plane using CASA's {\tt uvcontsub} routine. 
This method is suitable for our application because continuum emission dominates the source, and deconvolving the line emission will be more robust if it is not subject to deconvolution errors from the brighter continuum. 
However, continuum estimation in the $uv$-plane has the serious drawback that interpolating visibilities between channels is only an approximate solution for emission near the phase center, resulting in poorer performance for distributed emission farther from the phase center. 
This necessary accommodation does not imply a significant difference in the physical quantities derived in this study.
However, a more detailed treatment is developed for the H{\sc ii} regions in the Galactic plane sampled in the THOR survey extension to the Galactic center.

\begin{figure*}[h!]
\includegraphics[width=0.99\textwidth,angle=0,origin=c]{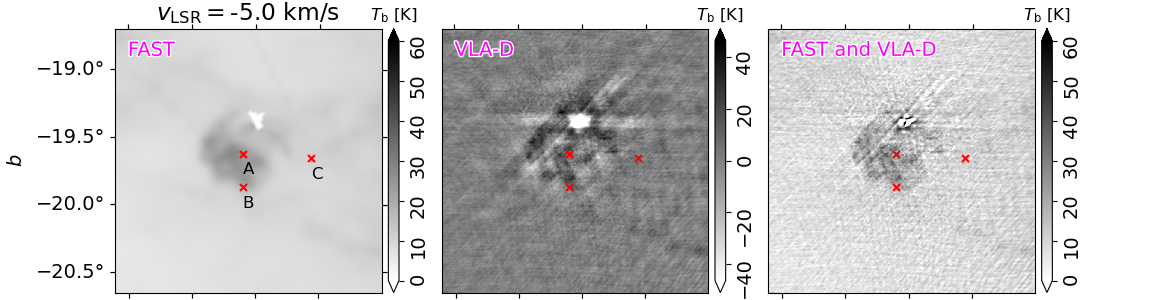}
\includegraphics[width=0.99\textwidth,angle=0,origin=c]{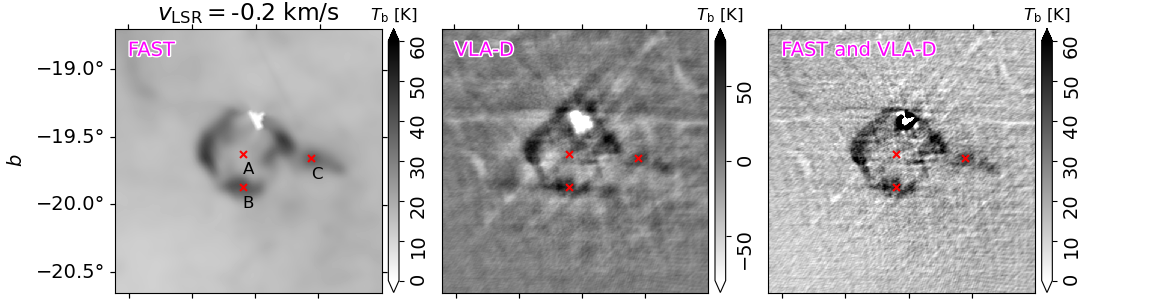}
\includegraphics[width=0.99\textwidth,angle=0,origin=c]{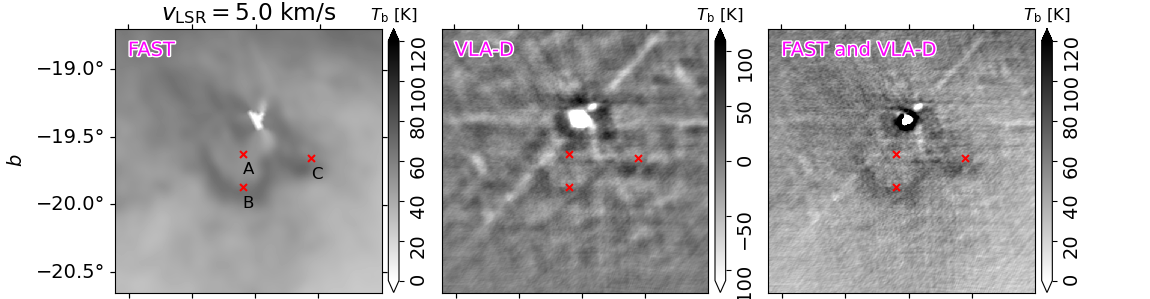}
\includegraphics[width=0.99\textwidth,angle=0,origin=c]{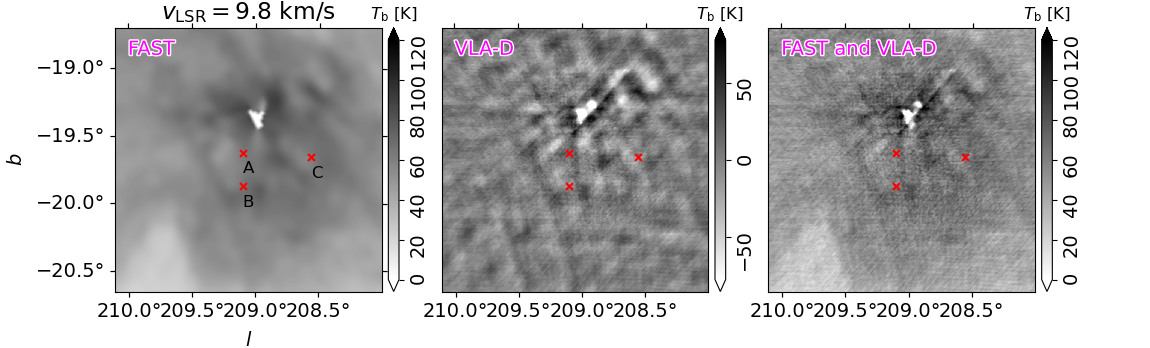}
\caption{H{\sc i} emission in selected velocity channels for the FAST, VLA D-array, and FAST and VLA D-array combined observations.
The crosses indicate the positions of the spectra presented in Fig.~\ref{fig:multiHIspectra}.
}
\label{fig:multiHI}
\end{figure*}

The imaging of the H{\sc i} was performed using CASA's {\tt tclean} routine.
An example of the VLA-D-only maps is presented in the central column of Fig.~\ref{fig:multiHI}.
The interferometric images reveal the EON's front hemisphere and its contours, as well as the unavoidable sidelobes produced by M42.

\begin{figure}[h!]
\centering{
\includegraphics[width=0.5\textwidth,angle=0,origin=c]{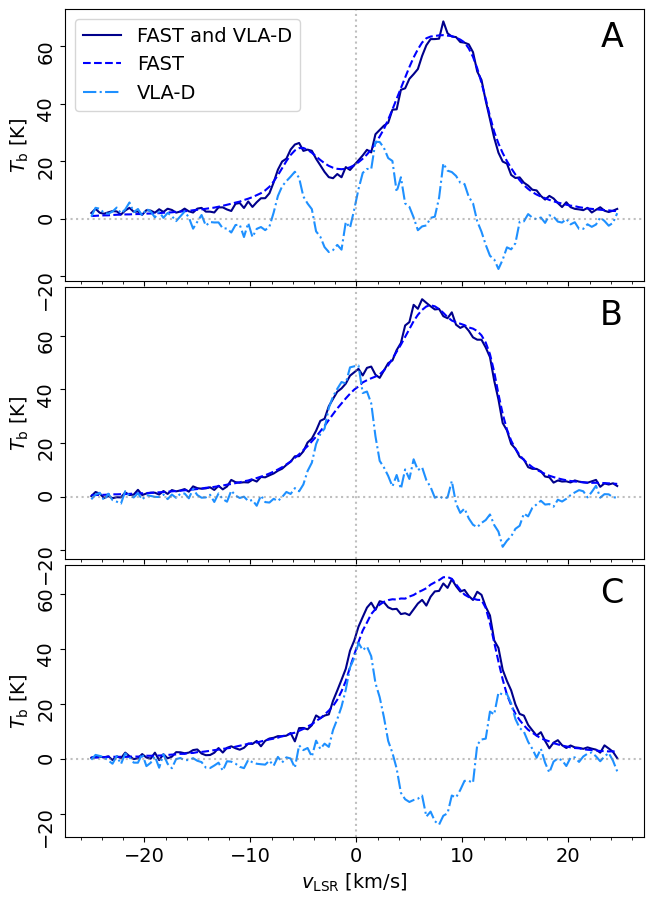}
}
\caption{H{\sc i} spectra for the positions indicated in Fig.~\ref{fig:multiHI}.}
\label{fig:multiHIspectra}
\end{figure}

\section{Combination with single-dish observations}\label{app:FAST}

We merged the FAST and VLA D-array observations using the {\tt feather} routine, which converts each image to gridded Fourier space, mixes them, and transforms the resulting combination back to real space.
We used the default {\tt feather} scale factor for the single-dish image ({\tt sdfactor}\,$=$\,1.0).
We do not apply spatial frequency filtering to the single-dish image ({\tt lowpassfiltersd}\,$=$\,{\tt False}).

We used as inputs the baseline-corrected CH{\sc i}NA observations projected into the VLA imaging grid. 
We aligned the CH{\sc i}NA maps with the interferometric images using the {\tt reproject} package and resampled into the same spectral axis using the {\tt spectral\_cube} in {\tt astropy}.
An example of the resulting H{\sc i} maps is presented in the left column of Fig.~\ref{fig:multiHI}.

Following the quality assessment metrics for interferometric images introduced in \cite{plunkett2023}, we evaluated the quality of our reconstruction using the fractional difference between the reconstructed and reference images:
\begin{equation}\label{eq:A}
A = \frac{R(v_{\rm LSR})-R_{0}(v_{\rm LSR})}{R_{0}(v_{\rm LSR})},
\end{equation}
where $R_{0}(v_{\rm LSR})$ is the reference image and $R(v_{\rm LSR})$ is the reconstructed image for the velocity channel centered on $v_{\rm LSR}$.
For the reference image, we use the FAST H{\sc i} map. 
The evaluated map is the combined VLA-D and FAST maps convolved to the angular resolution of the FAST data. 

\begin{figure}[h!]
\centering{
\includegraphics[width=0.4\textwidth,angle=0,origin=c]{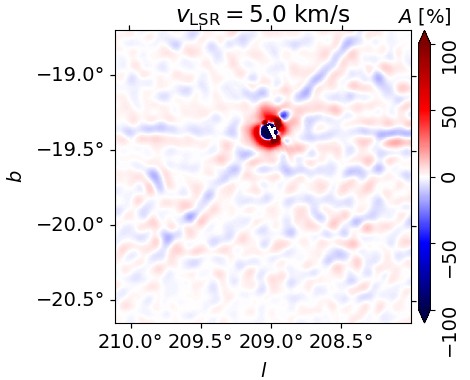}
}
\caption{Fractional difference between the FAST and combined FAST and VLA observations convolved to the FAST angular resolution for a reference velocity channel.}
\label{fig:Amap}
\end{figure}

Figure~\ref{fig:Amap} shows the distribution of the $A$-parameter for a reference $v_{\rm LSR}$\,$=$\,5\kps.
The largest variations are found toward the strong radio continuum source around the ONC, as expected from the lack of FAST data toward that region. 
These flux discrepancies extend up to 0\pdeg25 degrees around the center of that source.
There is also a significant difference toward M43, where the H{\sc i} absorption against the radio continuum sources is also not sampled by the FAST observations. 

\begin{figure}[h!]
\centering{
\includegraphics[width=0.4\textwidth,angle=0,origin=c]{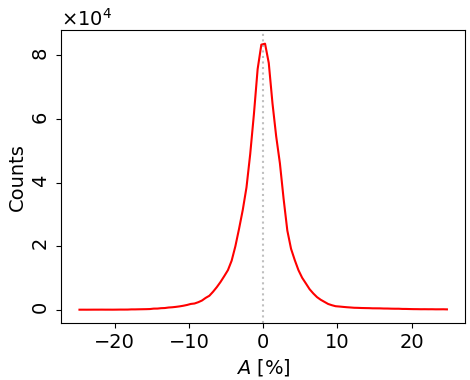}
}
\caption{Histogram of the fractional differences between the single-dish and combined observations, as defined in Eq.~\eqref{eq:A}.}
\label{fig:Ahist}
\end{figure}

Figure~\ref{fig:Amap} also reveals the pattern introduced by the VLA-D array sidelobes, which extend radially from the central position of the ONC.
Figure~\ref{fig:Ahist} shows that the amplitude of the fluctuations in the H{\sc i} map, excluding the ONC and M43 vicinities, is statistically small when compared to the whole extension of the maps, around 5\%.
This value serves as a reference for the uncertainty in the fluxes obtained from the interferometric reconstruction and data combination.

\begin{figure}[h!]
\centering{
\includegraphics[width=0.47\textwidth,angle=0,origin=c]{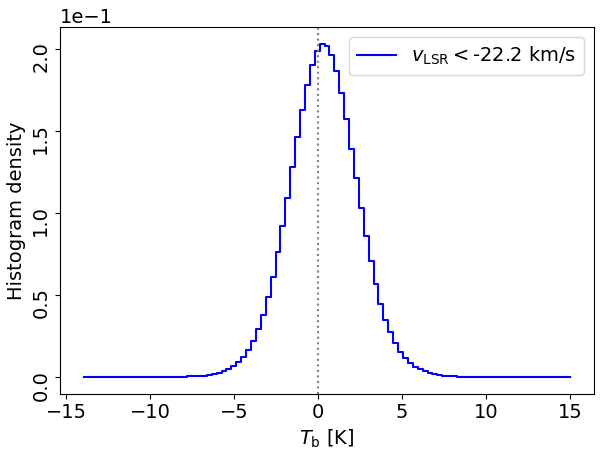}
}
\caption{Histogram of the $T_{\rm b}$ in the noise-dominated channels, which we employed to estimate in the combination of the VLA and FAST observations.}
\label{fig:noiseHist}
\end{figure}

We estimated the noise in the combined images by considering channels with low signal, which we identified in the range $v_{\rm LSR}$\,$<$\,$-22$\,\kps.
Figure~\ref{fig:noiseHist} shows a histogram of the emission in that range, from which we derive a root mean square (RMS) of around 2.04\,K.
We use that reference value for the \postlang{21 cm} line intensity uncertainties, but it does not account for interferometric reconstruction artifacts or the effects of the lack of H{\sc i} absorption in the FAST data. 

\section{NGC1977}\label{app:NGC1977}

\begin{figure}[h!]
\includegraphics[width=0.5\textwidth,angle=0,origin=c]{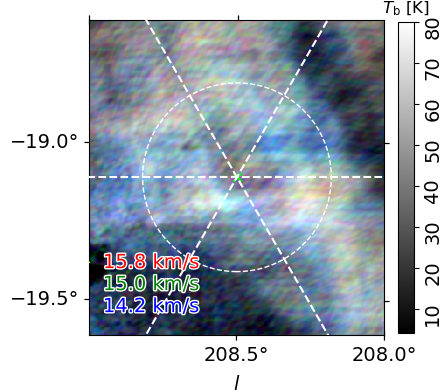}
\caption{Same as Fig.~\ref{fig:bubbleRGB}, but the position and central $v_{\rm LSR}$ of H{\sc ii} region NGC 1977.}
\label{fig:mapNGC1977}
\end{figure}

Our H{\sc i} observations also cover NGC 1977.
Although the shell around that H{\sc ii} region is not the subject of the main body of this paper, we show some generalities of the H{\sc i} emission toward that line of sight to facilitate subsequent studies.
Figure~\ref{fig:mapNGC1977} shows the emission across the velocity channels in which the cavity formed by the $\nu$ Ori star is most evident.

\begin{figure}[h!]
\includegraphics[width=0.5\textwidth,angle=0,origin=c]{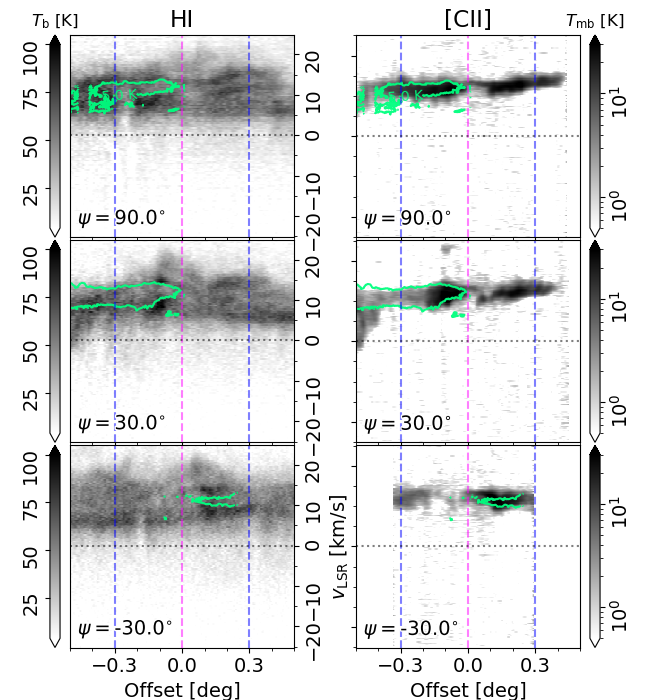}
\caption{Same as Fig.~\ref{fig:lvEONbubble}, but for the H{\sc ii} region NGC 1977 along the profiles shown on the middle panel of Fig.~\ref{fig:mapNGC1977}.}
\label{fig:pvNGC1977}
\end{figure}

\begin{figure}[h!]
\includegraphics[width=0.5\textwidth,angle=0,origin=c]{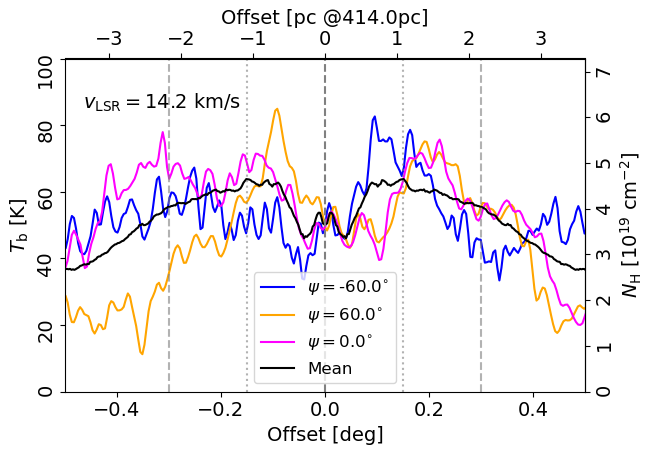}
\caption{Same as Fig.~\ref{fig:radialprofiles}, but for the H{\sc ii} region NGC 1977 along the profiles shown on the middle panel of Fig.~\ref{fig:mapNGC1977}.}
\label{fig:rpNGC1977}
\end{figure}

Figure~\ref{fig:pvNGC1977} shows the emission profiles across the lines marked in Fig.~\ref{fig:mapNGC1977}.
The H{\sc i} emission indicates a slight decrease in the position of the [C{\sc ii}] emission dent identified in \citetalias{pabst2020} as the NGC 1977 cavity.
However, the cavity is hard to identify in the H{\sc i} emission along.
The H{\sc i} emission radial profiles obtained at the $v_{\rm LSR}$ where apparent extension of the shell is maximum, shown in Fig.~\ref{fig:rpNGC1977}, indicate that the cavity has a radius $\sim$1\,pc.
The shell thickness is harder to define because it blends with the surrounding emission. 

\end{appendix}
\end{document}